\documentclass[iop,revtex4]{emulateapj}
\usepackage{color}
\usepackage{multirow}

\newcommand{\spt}{SPT-CL~J2106-5844}
\newcommand{\bvec}[1]{\textbf{#1}}
\newcommand{\LCDM}{$\Lambda$CDM~}
\newcommand{\HST}{{\it HST}}
\newcommand{\mytilde}{\raise.19ex\hbox{$\scriptstyle\sim$}}
\newcommand{\kms}{$\mbox{km}~\mbox{s}^{-1}$}

\newcommand{\solarm}{$10^{14}~M_{\sun}$}
\newcommand{\hsolarm}{$10^{14}h_{70}^{-1}~M_{\sun}$}
\newcommand{\persqarcmin}{arcmin$^{-2}$}
\newcommand{\sqarcmin}{arcmin$^{2}$}

\shorttitle{WL Study of SPT2106}
\shortauthors{Kim et al.}

\begin{document}

\title{Precise Mass Determination of SPT-CL~J2106-5844, 
 the Most Massive Cluster at $\lowercase{Z}>1$}

\author{Jinhyub Kim\altaffilmark{1, 2}, 
	M. James Jee\altaffilmark{1, 2},
	Saul Perlmutter\altaffilmark{3}, \\
	Brian Hayden\altaffilmark{4},
	David Rubin\altaffilmark{4},
	Xiaosheng Huang\altaffilmark{5},
	Greg Aldering\altaffilmark{3}, and 
	Jongwan Ko\altaffilmark{6,7}
	}

\affil{\altaffilmark{1}Department of Astronomy, Yonsei University, 50 Yonsei-ro, Seoul 03722, Korea; \texttt{jinhyub@yonsei.ac.kr; mkjee@yonsei.ac.kr}}
\affil{\altaffilmark{2}Department of Physics, University of California, Davis, One Shields Avenue, Davis, CA 95616, USA}
\affil{\altaffilmark{3}Physics Division, Lawrence Berkeley National Laboratory, 1 Cyclotron Road, Berkeley, CA, 94720, USA}
\affil{\altaffilmark{4}Space Telescope Science Institute, 3700 San Martin Drive, Baltimore, MD 21218, USA}
\affil{\altaffilmark{5}Department of Physics and Astronomy, University of San Francisco, 2130 Fulton Street, San Francisco, CA 94117, USA}
\affil{\altaffilmark{6}Korea Astronomy and Space Science Institute, 776, Daedeokdae-ro, Yuseong-gu, Daejeon 34055, Korea}
\affil{\altaffilmark{7}University of Science and Technology, Daejeon 34113, Korea}

\begin{abstract}
We present a detailed high-resolution weak-lensing (WL) study of \spt~at $z=1.132$, claimed to be the most massive system discovered at $z > 1$ in the South Pole Telescope Sunyaev-Zel'dovich (SPT-SZ) survey. 
Based on the deep imaging data from the Advanced Camera for Surveys and Wide Field Camera 3 on-board the {\it Hubble Space Telescope}, 
we find that the cluster mass distribution is asymmetric, composed of a main clump and a subclump $\mytilde640$~kpc west thereof.
The central clump is further resolved into two smaller northwestern and southeastern substructures separated by $\mytilde150$~kpc.
We show that this rather complex mass distribution is more consistent with the cluster galaxy distribution than a unimodal distribution as previously presented.
The northwestern substructure coincides with the BCG and X-ray peak while the southeastern one agrees with the location of the number density peak. These morphological features and the comparison with the X-ray emission suggest that the cluster might be a merging system. 
We estimate the virial mass of the cluster to be $M_{200c} = (10.4^{+3.3}_{-3.0}\pm1.0)~\times$~\solarm, where the second error bar is the systematic uncertainty. Our result confirms that the cluster \spt~is indeed the most massive cluster at $z>1$ known to date. We demonstrate the robustness of this mass estimate by performing a number of tests with different assumptions on the centroids, mass-concentration relations, and sample variance. 
\end{abstract}


\section{Introduction}
Careful studies of galaxy clusters, the largest gravitationally bound structures, play a pivotal role in understanding the large scale structure formation and evolution of the universe. 
The cluster mass function, the number of galaxy clusters at a given mass interval, enables us to probe the growth rate of the largest halos and constrain cosmological parameters \citep[e.g.,][]{Allen2011}.
Since the amplitude of the matter power spectrum probed by the cluster mass function at a fixed redshift is a degenerate function of the matter density $\Omega_m$ and normalization $\sigma_8$, it is necessary to combine the cluster mass functions over a wide range of redshift
in order to break the degeneracy. Therefore, there have been constant efforts to enlarge the sample of galaxy clusters at high redshift.

The efforts to find new high-$z$ clusters are happening at various wavelengths, which include 
the X-ray (e.g., \citealt{Fassbender2011}; \citealt{Mehrtens2012}), 
optical/IR (e.g., \citealt{Muzzin2009}; \citealt{Adami2010}), 
and millimeter (e.g., \citealt{Marriage2011}; \citealt{Bleem2015}) surveys. 
Among these, the millimeter surveys utilizing the Sunyaev-Zel'dovich (SZ) effect are particularly efficient in detecting massive high-$z$ (e.g., $z>1$) objects because the SZ signal strength is nearly independent of the cluster redshift without being plagued by the redshift-dependent surface brightness dimming $(1+z)^{-4}$.

The galaxy cluster, \spt~(hereafter SPT2106) was discovered in the South Pole Telescope (SPT) SZ survey with a high signal-to-noise (18.5$\sigma$ at 150 GHz) detection (\citealt{Foley2011}, hereafter F11). 
F11 performed follow-up observations including both photometry and spectroscopy at various wavelengths. 
Their infrared and X-ray observations from the {\it Spitzer} and {\it Chandra} space telescopes, respectively, show that the cluster is also rich in galaxy and intracluster medium. 
Both its high X-ray temperature $T_{X} = 11.0^{+2.6}_{-1.9}~\mbox{keV}$ and luminosity $L_{X}(0.5-2~\mbox{keV}) = (13.9\pm1.0)\times 10^{44}~\mbox{erg}~\mbox{s}^{-1}$ are strong indications that the system is very massive, although the morphology of the X-ray emission suggests that the system might be undergoing a merger and thus the temperature-based mass measurement might have been somewhat overestimated. 
F11 confirmed 18 spectroscopic member galaxies (mostly early-type) and measured the velocity dispersion to be $1230^{+270}_{-180}$~\kms. This high velocity dispersion is in support of the cluster's extreme mass.

F11 derived the virial mass of SPT2106 $M_{200c}=(12.7\pm2.1)~\times$~\hsolarm, combining their SZ, X-ray, and velocity dispersion measurements. This mass is the highest among the entire $z>1$ sample of the SPT-SZ survey \citep{Reichardt2013}. Follow-up studies with SZ (e.g., \citealt{Williamson2011}; \citealt{Reichardt2013}; \citealt{Bleem2015}) and X-ray (e.g., \citealt{Amodeo2016}; \citealt{Bartalucci2017}) observations consistently provided high masses for the system. \cite{Schrabback2018} (hereafter S18) presented the first weak-lensing (WL) measurement of SPT2106 using the Advanced Camera for Surveys (ACS) in the {\it Hubble Space Telescope} (\HST) data and obtained $M_{200c}=8.8^{+5.0}_{-4.6}~\times$~\solarm. 
The central value of this measurement is somewhat lower than the previous non-WL estimates (e.g., $\mytilde30$\% smaller than the F11 value). However, because of the large ($\gtrsim50$\%) error bars on the S18 WL result, the statistical significance of the difference is not strong.

The abundance of massive clusters at high redshift such as SPT2106 is a sensitive function of cosmological parameters. 
Hence, the presence of any exceptional cluster can create a non-negligible tension with the current \LCDM paradigm. According to F11, the expected number of SPT2106-like clusters is estimated to be $\mytilde1$ ($\mytilde0.07$) in the entire sky (parent 2500 sq. deg survey). Currently, the largest source of error in the abundance estimation is the mass estimate uncertainty.

In this paper, we present a detailed WL analysis of SPT2106 using ACS and Wide Field Camera 3 (WFC3) data. Our work is different from S18 as follows. First, we measure WL signals from both ACS and WFC3 images while S18 obtained galaxy shapes only from ACS. As will be discussed in \S\ref{section_analysis}, high-redshift galaxies are brighter in IR, which enables us to obtain more precise shapes and a higher (a factor of $\mytilde9$ in the current case) number density of sources.
Second, our WL pipeline uses the ``SFIT" method, whose shear calibration accuracy has been publicly validated in the most recent public data challenge called the third GRavitational lEnsing Accuracy Testing \citep[GREAT3;][]{GREAT3}. SFIT won eight GREAT3 branches, one of which is the analysis of simulated future space-based WL images containing various effects such as complex variation on point spread function (PSF), multi-epoch dithering, point-spread-function undersampling, real galaxy morphologies, etc. 
The S18 analysis is based on the \cite{KSB1995} method, which has been improved by a number of authors (e.g., \citealt{Hoekstra1998}; \citealt{Erben2001}; \citealt{Schrabback2010}). Given the importance and rarity of the target, it is useful to study SPT2106 with an independent pipeline.
Third, we investigate WL substructures of SPT2106. Our source density is more than a factor of 9 higher than the one in S18. This increase is due to the difference in the WL pipeline, source selection method, and the availability of the deep WFC3 imaging data. Our higher source density enables us to probe the substructure in greater detail.

This paper is organized as follows. We describe our \HST~data and their reduction in \S\ref{section_observations}. 
In \S\ref{section_analysis}, we present our WL analysis procedure. Our mass reconstruction and virial mass estimates are shown in \S\ref{section_results}.
We compare our WL mass estimates with other mass proxies and previous WL measurement, discuss possible sources of systematic uncertainties, and estimate the rarity of such a massive cluster in \S\ref{section_discussion}.
We summarize our work and present conclusions in \S\ref{section_summary}.

Throughout the paper, $M_{200c}$ (or $M_{500c}$) corresponds to the mass enclosed within a radius, inside which the mean density equals 200 (500) times the critical density of the universe at the cluster redshift.
We adopt the cosmology published in \cite{Planck2016}. 
For this cosmology the angular diameter distance to the cluster is $\mytilde1741$~Mpc, and thus, the plate scale is $\mytilde506$~kpc~$\mbox{arcmin}^{-1}$ at the cluster redshift.
We use the AB magnitude system corrected for the Milky Way foreground extinction and express all uncertainties as the $1\sigma$ confidence ($\mytilde68.3$\%) level.

\begin{figure}
\centering
\includegraphics[width=8cm]{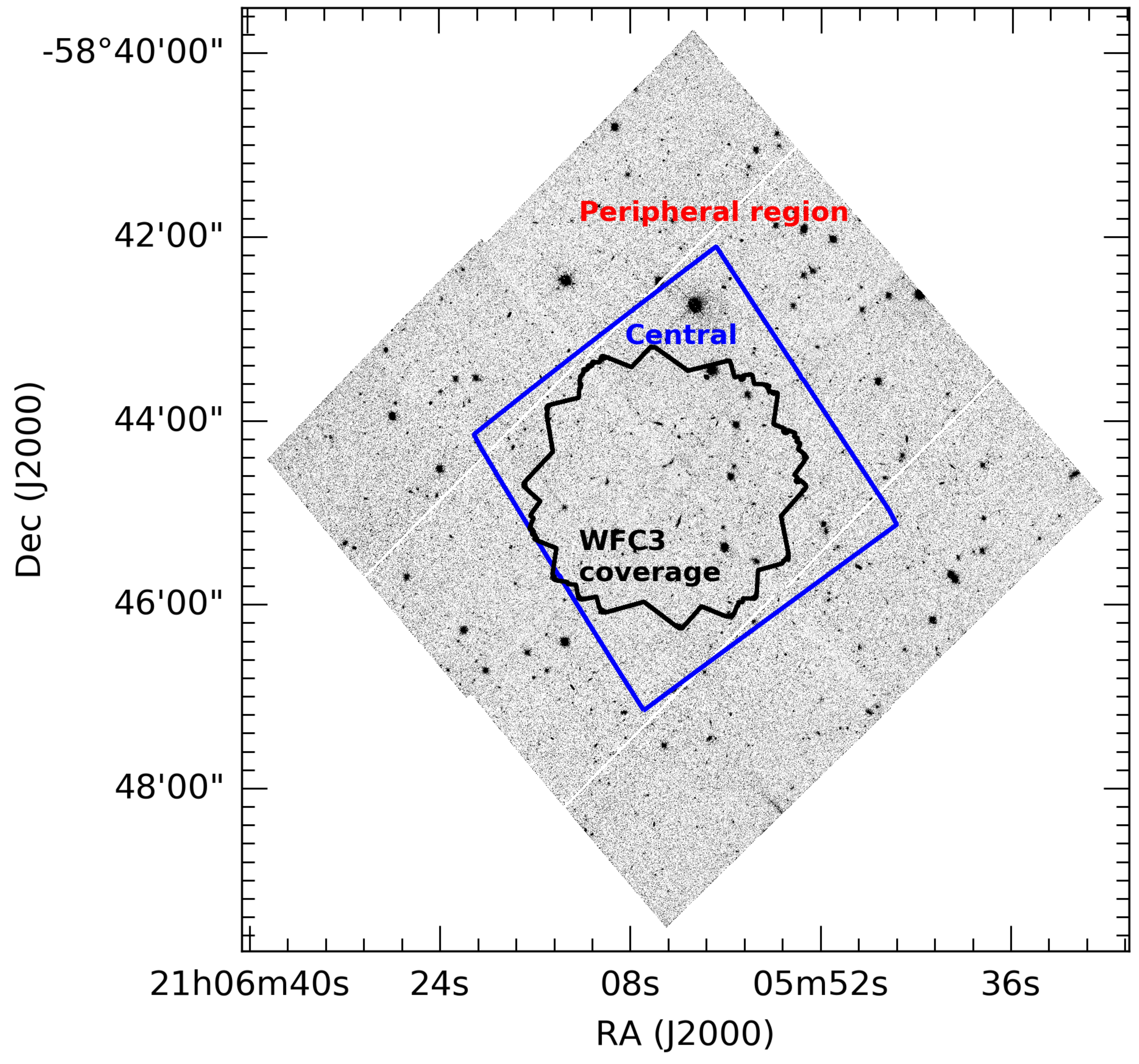}
\caption{\HST~observation footprint of SPT2106. 
In inverted gray scale,
we show the intensities of ACS/F606W, which observed the cluster in the $2\times2$ mosaic pattern and covered the $\mytilde6\times6$ arcmin$^2$ region.
The blue box depicts the central ($\mytilde3\times3$ arcmin$^2$) F814W pointing. 
Throughout the paper, we refer to the areas inside and outside this pointing as
central and peripheral regions, respectively. 
The WFC3/IR pointings (black) lie within the central region.
}
\label{fig:SPT2106field}
\end{figure}

\section{Observations} \label{section_observations}
In this study we use two \HST~observation programs for the galaxy cluster SPT2106.
In PROP 12477 (Cycle 19, PI: W. High), the cluster was observed in 2011 October and 2012 December with the Wide Field Channel (WFC) of the ACS using the F606W and F814W filters. 
The F606W observation used a 2 by 2 mosaic pattern whereas the F814W filter observed the central region in a single pointing. 
The total integration times per pointing are 1,920 s and 1,916 s for F606W and F814W, respectively.
PROP 13677 (Cycle 22, PI: S. Perlmutter) observed the cluster in a single pointing with the WFC3/UVIS F814W, WFC3/IR F105W, and WFC3/IR F140W from 2015 March to 2015 September with total integration times of 3,491 s, 7,181 s, and 8,131 s, respectively. 
Hereafter, we refer to the region covered by ACS/F814W as the ``central" region and the area outside this region as the ``peripheral" region as illustrated in Figure~\ref{fig:SPT2106field}. Note that WFC3/IR covers a slightly smaller region than the central region. Because object colors are available only in the central region, we use different schemes for source selection and redshift estimation for the two regions (\S\ref{section_beta}).

We start our data reduction using the {\tt FLC} and {\tt FLT} images provided by the STScI CALACS pipeline \citep{CALACS} for the ACS and WFC3/IR data, respectively.
The {\tt FLC} images have been corrected for charge transfer efficiency (CTE) degradation based on the \cite{Ubeda2012} algorithm. 
The CTE degradation is mainly due to radiation damage and, prior to correction, manifests itself as a long trail along the CCD parallel readout direction. 
Because the success of WL analysis depends on our ability to control systematics that affect galaxy shapes, it is crucial to verify that the STScI pipeline CTE correction is sufficiently accurate. 
In \cite{Jee2014}, we quantified the accuracy by utilizing the shapes of cosmic rays and demonstrated that the residual error in the STScI pixel-based correction is sufficiently small, compared to the statistical noise in cluster lensing analysis, although it is shown that the imperfection of the method slightly overcompensates for the CTE degradation for very faint objects.
Contrary to the ACS/WFC and WFC3/UVIS detectors, the WFC3/IR detector does not suffer from CTE degradation. 

To align individual exposures, 
we select one of the ACS/F814W exposures as a reference image and estimate shifts for the rest of the exposures with respect to this reference image based on common astronomical sources. 
This explicit shift estimation is mandatory because the typical WCS information in \HST~image headers is inaccurate.
We use the MultiDrizzle \citep{2002multidrizzle} software in order to conduct cosmic ray removal, sky subtraction, and image combination. We adopt the Lanczos3 drizzle kernel for ACS/WFC and WFC3/UVIS with the output pixel scale of 0\farcs05~per pixel. 
Our previous tests have shown that for ACS/WFC this combination of drizzle parameters produces the sharpest PSF \citep{Jee2007}. 
Because the WFC3/IR PSF is severely undersampled,
we cannot use the same Lanczos3 drizzling kernel to process the image. Instead, we use the Gaussian kernel while maintaining the same 0\farcs05~pixel scale.

We detect objects with SExtractor \citep{Bertin1996} in dual-image mode (using one image for detection and the other for measurement). 
The detection image is created by weight-averaging all available filter images. 
We find sources by looking for at least 10 connected pixels above 1.5 times the sky rms. The total number of sources obtained is $\mytilde9,800$ within the $\mytilde6\arcmin\times6\arcmin$. 
We exclude spurious objects (e.g., diffraction spikes of bright stars, cosmic rays, star formation regions in large foreground galaxies, etc.) with visual inspection.

\section{Analysis} \label{section_analysis}
We measure galaxy shapes in ACS/F606W and WFC3/F140W. As mentioned in \S\ref{section_observations}, the F606W filter covers the $6\arcmin\times6\arcmin$ cluster field whereas the WFC3/F140W filter is used only for the observation of the central $2\arcmin\times2\arcmin$ region. We combine the two shape catalogs where available. 
We must adopt slightly different source selection and redshift estimation schemes between the central and peripheral regions due to the difference in color availability. 
In \S\ref{section_PSF} and \S\ref{section_shape_measurement}, we discuss our PSF modeling and shape measurement, respectively. Source selection and redshift estimation schemes are described in \S\ref{section_source_selection} and \S\ref{section_beta}, respectively.

\subsection{Point Spread Function Modeling} \label{section_PSF}
Modeling accurate PSFs is one of our supreme interests because a PSF dilutes lensing signals and induces artificial ellipticity correlations.
The effect is more destructive for small, faint galaxies, which provide more efficient lensing signals than large, bright galaxies because they are on average at higher redshift. Thus, accurate PSF modeling is paramount when one studies high-redshift clusters.

In \HST, it is well-known that the PSF changes in time because of the so-called ``focus breathing" effect and in position across the CCD chip because of the position-dependent optical aberration.
Since there are only several high S/N stars available within each pointing, it is impossible to obtain a high-quality PSF model from the image itself that represents the complicated spatial variation. However, thanks to the repeatability of the pattern in the PSF variation \citep{Jee2007}, one can obtain a high-fidelity PSF model from dense stellar fields and apply it to a WL image. 
This PSF library approach has been successfully applied to a number of clusters in our and other previous studies (e.g., \citealt{Jee2005}; \citealt{Schrabback2010}). In the current study, we continue to use this library-based method to model PSFs and refer readers to our previous papers (e.g., \citealt{JeeTyson2009}) for details.

\subsection{Shape Measurement} \label{section_shape_measurement}
The distortion of the shape of a background source due to a foreground lens is expressed by the matrix below:
\begin{equation}
\textbf{A}=(1-\kappa) \left (\begin{array} {c c} 1 - g _1 & -g _2 \\
                      -g_2 & 1+g _1
          \end{array}  \right ), \label{eqn_lens_trans}
\end{equation}
\noindent
where the convergence $\kappa$ is the projected mass density in the unit of the critical surface density $\Sigma_c$ and the reduced shear $g_{1(2)}$ determines the amount of stretch along the $x$-axis (along the 45\degr~direction). 
The critical surface density $\Sigma_c$ is defined by
\begin{equation}
\Sigma_c = \frac{c^2}{4 \pi G} \frac{D_s}{D_l D_{ls} }, \label{eqn_sigma_c}
\end{equation}
\noindent
where $c$ is the speed of light, $G$ is the gravitational constant. $D_l$, $D_s$, and $D_{ls}$ are the angular diameter distance to the lens, to the sources, and between the lens and the sources, respectively. The ratio of the last two distances $\beta=D_{ls}$/$D_s$ is often referred to as ``lensing efficiency" and is estimated from the effective redshift of the source plane (\S\ref{section_beta}).

Observationally, the reduced shear $g$ must be inferred from a population of source galaxies because the amount of distortion in each galaxy image is very small compared to its intrinsic shape dispersion. Assuming that the intrinsic galaxy shape is random, we can estimate the reduced shear $g$ by averaging ellipticities $g=\left < e \right >$. 
We define the ellipticity of a galaxy image to be $e = (a-b)/(a+b)$, where $a$ and $b$ are the semi-major and -minor axes, respectively. In practice, the relation $g=\left < e \right >$ does not hold exactly due to a number of reasons. 
One obvious reason is that there is no unique way to define the semi-major and minor axes of real galaxy images. Most galaxy images possess radially varying isophotes, which are also asymmetric and irregular.
Therefore, one's definition of ellipticity must be only operational and we need to derive a correction factor that reconciles the difference between $g$ and $\left <e \right>$.

In this paper, we measure ellipticity by fitting a two-dimensional elliptical Gaussian to a galaxy image using the {\tt MPFIT} \citep{MPFIT} package. 
The elliptical Gaussian model is convolved with a model PSF computed for each object prior to fitting. Although formally the total number of free parameters is seven: 2 for the object centroid, 2 for the normalization and background level\footnote{The background level that we used is a local SExtractor background value.}, 2 for the semi-major and -minor axes, and 1 for the orientation angle, we fix the background level and centroid using the SExtractor output. 
The semi-major $a$ and -minor axes $b$ and the orientation $\phi$ are converted to
the galaxy ellipticities $e_1$ and $e_2$ as follows:
\begin{eqnarray}
e_1&=&e \cos{2 \phi} \\
e_2&=&e \sin{2 \phi}.
\end{eqnarray}
\noindent

As mentioned above, averaging the ellipticity obtained in this way causes
bias in our reduced shear $g$ estimation. The bias due to the difference between elliptical Gaussian and real galaxy profiles is referred to as ``model bias." And the bias caused by nonlinear translation of pixel noise to ellipticity is termed ``noise bias." (\citealt{Melchior2012}; \citealt{Refregier2012}). 
In addition to ``model bias" and ``noise bias," there are many
other sources of bias such as selection effect, blending, etc.

Instead of addressing individual sources of bias separately,
we ran WL image simulations to determine the combined effect. 
Our WL image simulations are described in \cite{JeeTyson2011} and \cite{Jee2013}. 
\cite{JeeTyson2011} explain in detail how we sampled galaxy images from the Hubble Ultra Deep Field (HUDF; \citealt{UDF06}) data. \cite{Jee2013} describe how shear calibrations are derived by comparing input and output (recovered) shears.
The original simulation tool was developed mainly for ground-based imaging data.
For {\it HST} WL analysis, an important change must be made because the PSF size is significantly different between ground and space.
Previously, when shearing galaxy images, we ignored the fact that the images had been convolved by the HST PSF, which is much smaller than those for LSST and the Deep Lens Survey \citep{Jee2013}.
However, this approximation is not valid in {\it HST} WL studies.
Therefore, we first deconvolved the HUDF galaxy images and then applied shears. Finally, we reconvolved the shear image with the ACS PSF, following the method described in the GalSim package \citep{Galsim}. 
The multiplicative factor obtained from this simulation is 1.15 for ACS/F606W. 

For WFC3/F140W images, there are some additional sources of systematics due to the detector characteristics such as undersampling, nonlinearity, interpixel capacitance, etc. 
Instead of including these features in our image simulation, for WFC3/F140W we chose to derive shear calibration utilizing $\mytilde2700$ common astronomical objects between ACS/WFC and WFC3/F140W in the archival data. We refer readers to \cite{Jee2017} for details. The resulting multiplicative factor for WFC3/F140W is 1.25. 
We combine the ACS/F606W and WFC3/F140W ellipticities if available by weight-averaging.

\begin{figure}
\centering
\includegraphics[width=8cm]{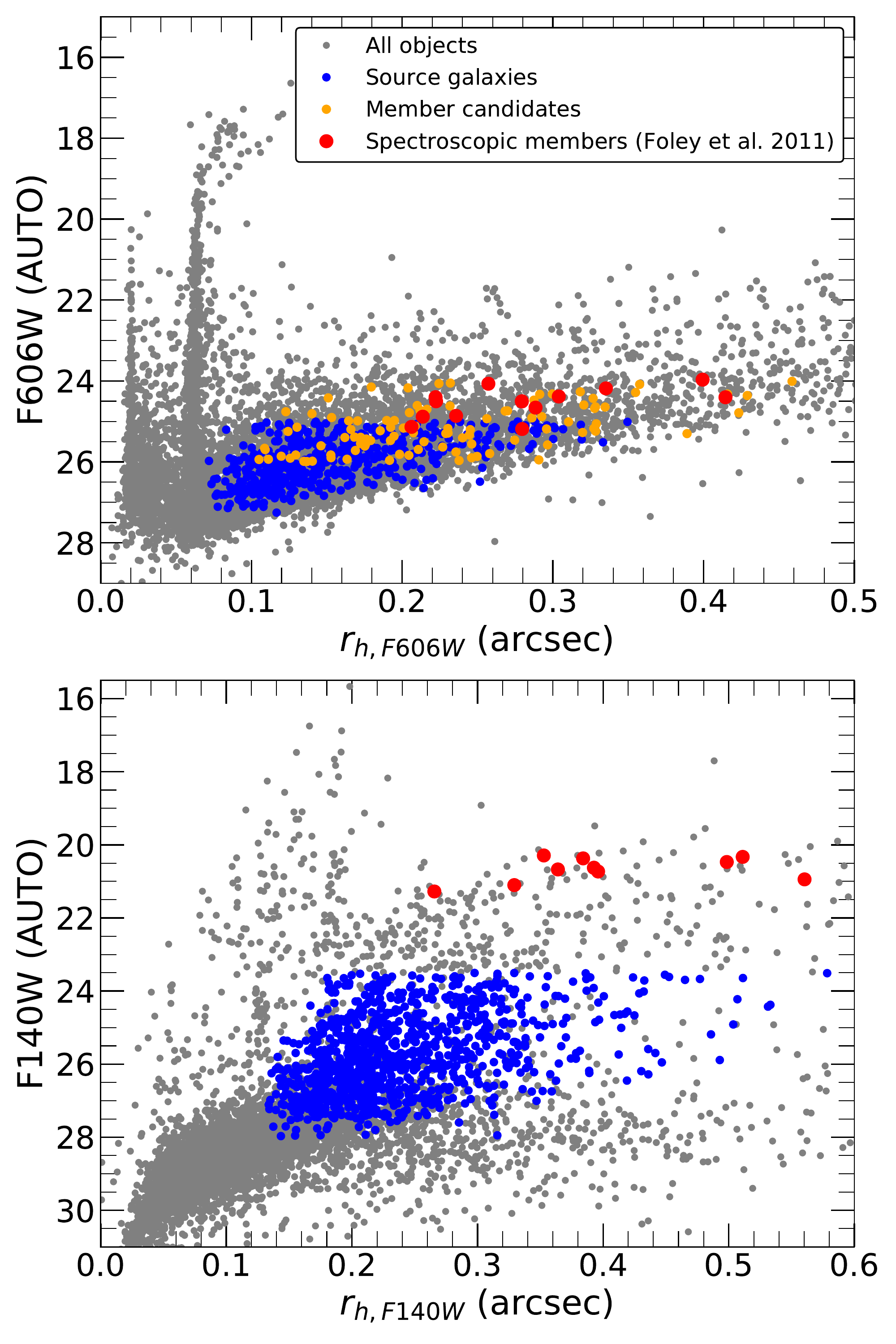}
\caption{Magnitude versus half-light radius relation. 
We show the relations obtained from the F606W and F140W data in the top and bottom panels, respectively.
We identify stars using their tight size-magnitude relation 
($r_{h}\sim$~0\farcs07~and $\sim$~0\farcs13~for F606W and F140W, respectively). 
Also displayed are our sources (blue), confirmed cluster members from optical spectroscopy (red), and member candidates (orange). Our sources are selected from both shape and photometric requirements (see text). 
The member candidates are selected assuming the presence of a 4000~\AA~break feature, and the effect of this on the magnitudes.
}
\label{fig:r50_m606}
\end{figure}

\begin{figure}
\centering
\includegraphics[width=8cm]{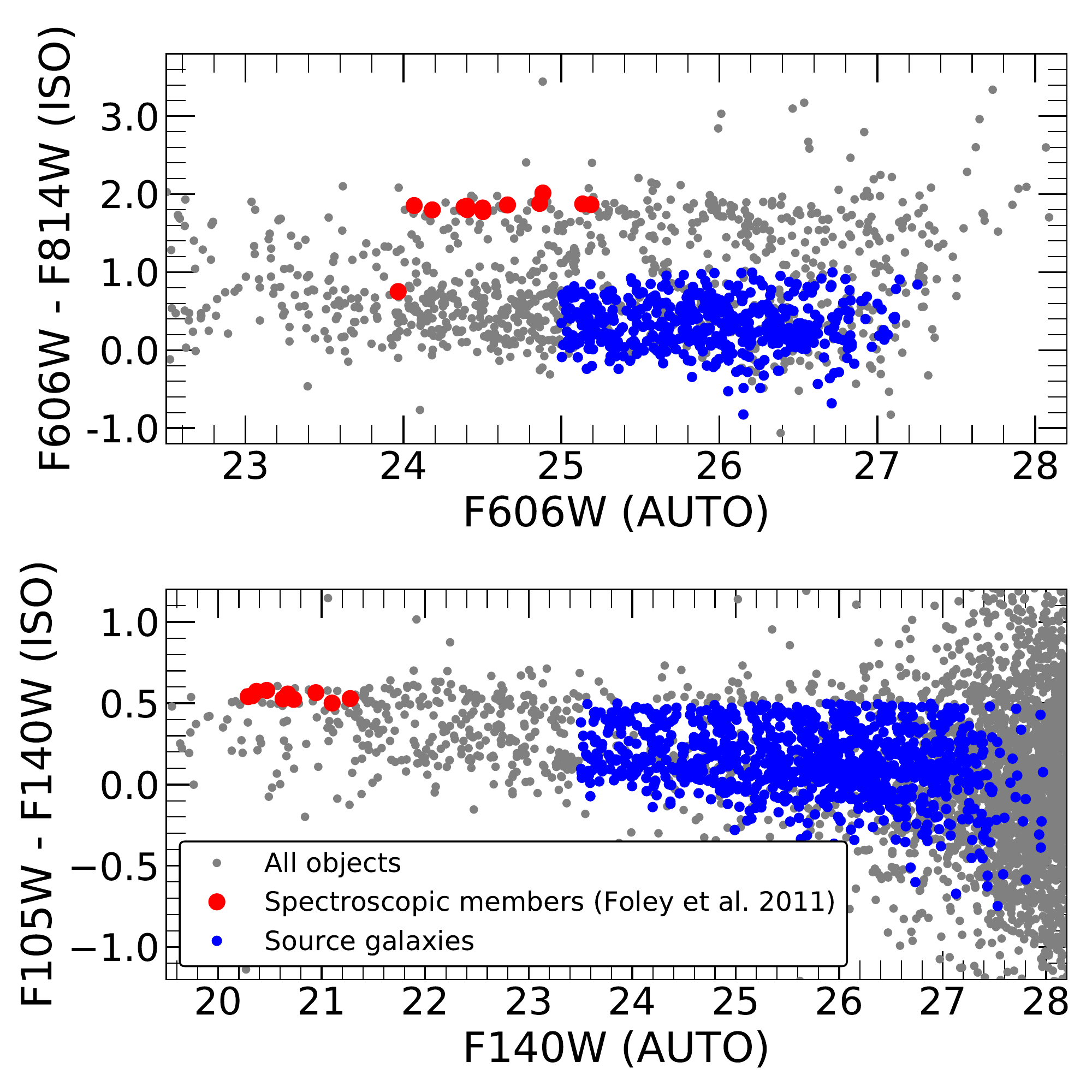}
\caption{Color-magnitude relation of the objects in the central SPT2106 field. The top (bottom) panel shows the result 
for the central $\mytilde3\times3$~\sqarcmin ~($\mytilde2\times2$~\sqarcmin) region seen in $\mbox{F606W}-\mbox{F814W}$ versus $\mbox{F606W}$ ($\mbox{F105W}-\mbox{F140W}$ versus $\mbox{F140W}$). 
Sources (blue circles) are selected based on these color-magnitude relations (top: $\mbox{F606W}-\mbox{F814W}< 1.0$, $25.0 < \mbox{F606W} < 28.0$, bottom: $\mbox{F105W}-\mbox{F140W}< 0.5$, $23.5 < \mbox{F140W} < 28.0$) and the shape selection criteria (see text). 
The spectroscopically confirmed cluster galaxies from \cite{Foley2011} are depicted in red. The confirmed members are well isolated in $\mbox{F606W}-\mbox{F814W}$ color, which brackets the rest-frame 4000~\AA~break feature. The object at F606W~$\sim24$ with $\mbox{F606W}-\mbox{F814W}\sim0.7$ is the late-type member confirmed by its [\ion{O}{2}] line \citep{Foley2011}. 
Interestingly, the red members occupy the narrow locus also in the $\mbox{F105W}-\mbox{F140W}$ color, although the two filters are not optimal in capturing their rest-frame 4000~\AA~break feature. 
}
\label{fig:cmd}
\end{figure}

\begin{figure}
\centering
\includegraphics[width=8cm]{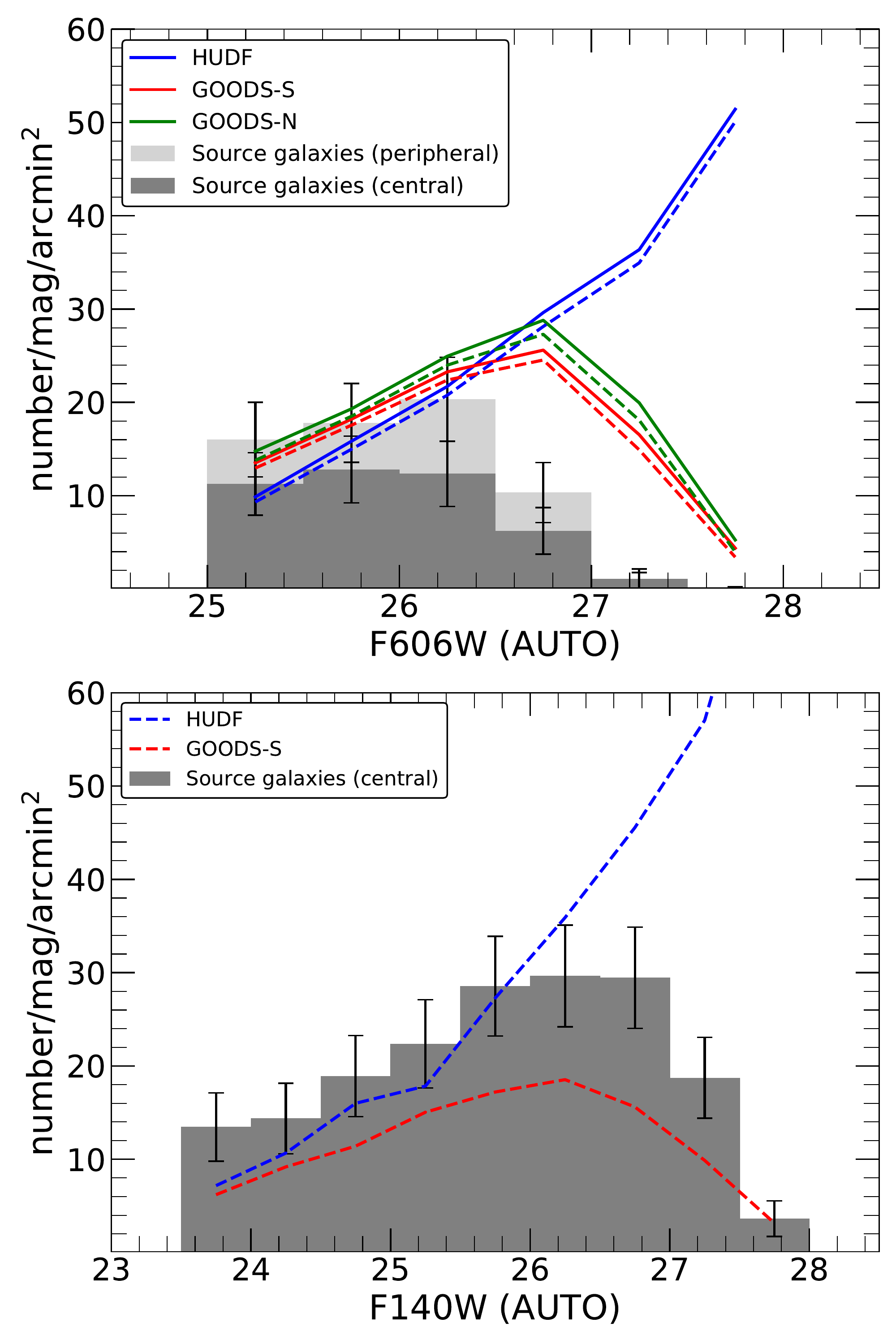}
\caption{Magnitude distribution of source galaxies.
Top: F606W magnitude comparison between SPT2106~and the three control fields: HUDF, GOODS-S, and GOODS-N. 
The light gray histogram shows the distribution in the peripheral region where only F606W is available (i.e., no color information). The dark gray histogram is obtained for the central region where we use the $\mbox{F606W}-\mbox{F814W}$ color to remove the cluster red sequence. Error bars are computed using Poissonian statistics. 
Solid lines represent the control field results when we apply only the magnitude selection criterion whereas dashed lines show the results when we use both object magnitudes and colors.
We interpret the consistency between the dark gray histogram and the dashed lines as indicating that the blue member contamination is not significant. Note that the large discrepancies at F606W $> 26.5$ result from the difference in depth.
See the text for the description of our correction for the red galaxy contamination in the peripherical region. \\
Bottom: Comparison of the F140W magnitude distribution between SPT2106~and the two control fields: HUDF and GOODS-S. 
The source selection based on $\mbox{F105W}-\mbox{F140W}$ provides a consistent magnitude distribution with the HUDF one.
}
\label{fig:mag_distribution}
\end{figure}

\begin{figure}
\centering
\includegraphics[width=8cm]{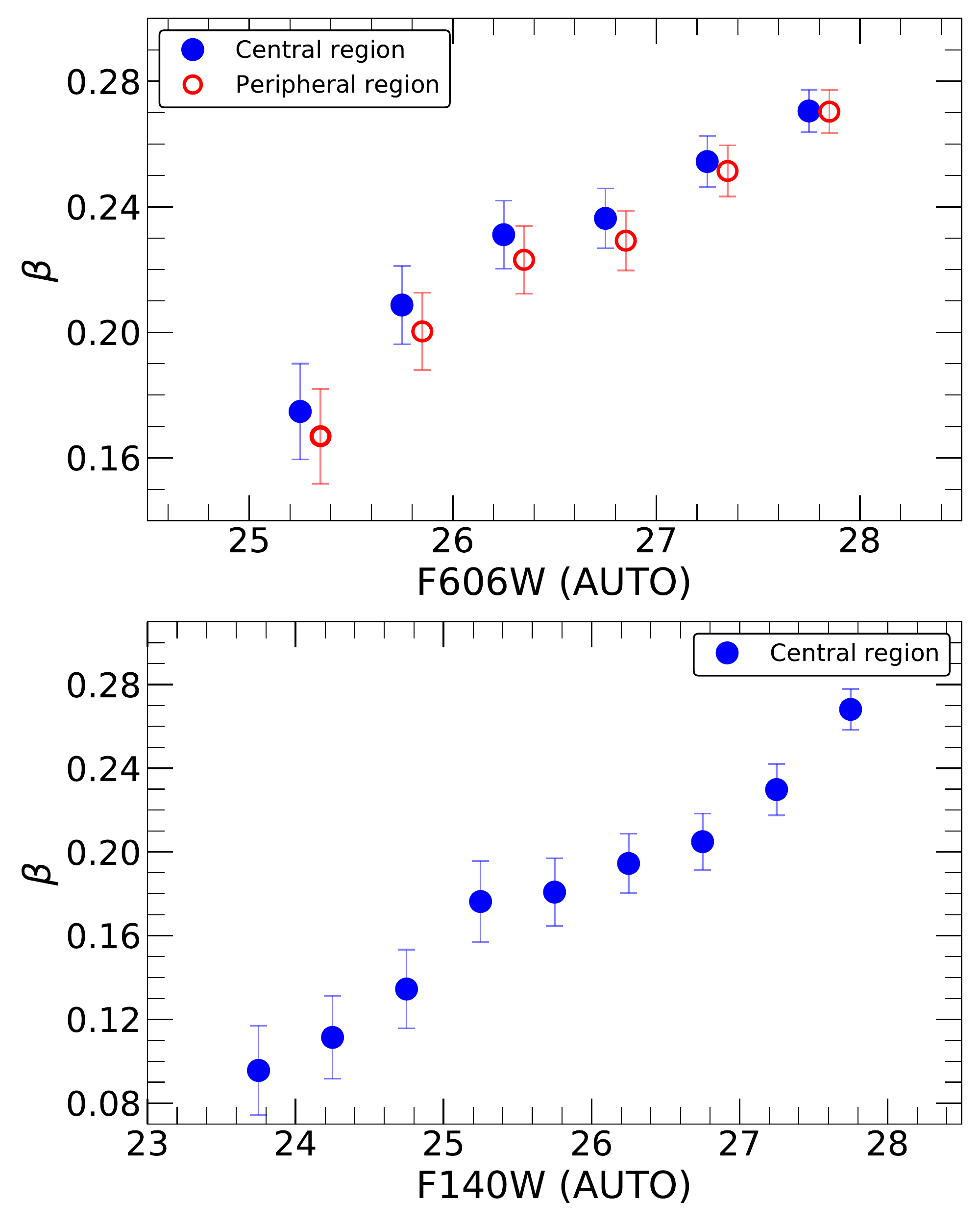}
\caption{Lensing efficiency ($\beta$) estimated for source population as a function of magnitude using HUDF as a control field for ACS (Top) and WFC3 (Bottom).
The lensing efficiency increases with magnitude, which is in accordance with our expectation that fainter galaxies are at higher redshifts.
In the central region, we use both object colors and magnitudes to select sources (blue filled circles) whereas
in the peripheral region (open red circles) we only apply magnitude cuts. We apply the same criteria to objects in HUDF.
See text for the description on how we use this color-magnitude-redshift or magnitude-redshift relation to derive the source redshift distribution after taking into account various effects such depth difference, red galaxy contamination, etc. 
}
\label{fig:mean_beta}
\end{figure}

\subsection{Source Selection} \label{section_source_selection}
We select the background sources that satisfy both shape and photometric requirements described in \S\ref{shape_requirements} and \S\ref{phot_requirements}, respectively. 
We impose different magnitude and color cuts on ACS and WFC3 instruments. 
Based on these requirements a total of 3129 objects are classified as sources. 
The resulting source density ($\mytilde79$~\persqarcmin)~is higher than the one in S18 by nearly a factor of 9. 
We discuss our evaluation of the lensing efficiency $\beta$ in \S\ref{section_beta}.

\subsubsection{Shape Requirement} \label{shape_requirements}
Our WL sources are selected based on the following shape criteria. 
First of all, ellipticities of sources (after PSF deconvolution) should have converging values. We fulfill this condition by selecting sources whose {\tt MPFIT} {\tt STATUS} parameter \citep{MPFIT} is unity. 
Second, objects should have their semi-minor axis from our elliptical Gaussian fitting (after PSF correction) greater than 0.4 pixels, which efficiently excludes point sources and small galaxies whose shapes are uncertain and subject to residual PSF systematics.
Third, errors on ellipticity should be less than $0.25$. 
The objects that have ellipticity error larger than this tend to have bright neighbors, very low surface brightness, and/or very small FWHM values. Moreover, their noise bias is large, which makes their shear calibration highly uncertain. 
Lastly, half-light radii (before PSF deconvolution) of the objects should be larger than those of stars (Figure~\ref{fig:r50_m606}). 
This requirement also prevents us from including spurious features such as cosmic rays, hot pixels, etc. 
In spite of the above efforts, some spurious objects (e.g., diffraction spikes around bright stars, cosmic rays, fragmented parts of foreground galaxies, clipped sources at the field boundaries, etc.) still remain. We manually identify and remove these objects by visual inspection.

\subsubsection{Photometric Requirement} \label{phot_requirements}

{\bf Source selection with ACS photometry.}
The redshifted rest-frame 4000~\AA~break feature (\mytilde8500~\AA~at the cluster redshift) at $z=1.132$ is bracketed by F606W and F814W.
The early-type members occupy a narrow locus in the color-magnitude diagram shown in the top panel of Figure~\ref{fig:cmd}. 
We select objects bluer and fainter than the red-sequence ($\mbox{F606W}-\mbox{F814W}< 1.0$ and $\mbox{F606W}>25$) as our background sources (blue dots). This enables us to remove only red cluster galaxies and thus the resulting catalog may include blue cluster members. 
In the peripheral region where only F606W is available, we select sources purely based on their magnitudes $25.0 < \mbox{F606W} < 28.0$. Therefore, we need to correct for the contamination for both red and blue members.

In order to assess the cluster member contamination, we utilize two control fields: the HUDF~\citep{UVUDF} and the Great Observatories Origins Deep Survey (GOODS; \citealt{GOODSmag}).
After applying the same selection criteria to the control fields, we compare the resulting magnitude distributions with those from our sources as shown in the top panel of Figure~\ref{fig:mag_distribution}.
If the contamination is significant in our source selection, it should appear as a significant number density {\it excess} beyond the sample variance.

In the central region (blue box in Figure~\ref{fig:SPT2106field}) where the $\mbox{F606W}-\mbox{F814W}$ color is available, the magnitude distribution (dark gray) is consistent with those from the control fields at the $\mbox{F606W}\lesssim26$ regime. The small difference between GOODS-N and GOODS-S suggests that the sample variance might be much smaller than the Poissonian scatter, shown with error bars.
For fainter sources, the control field densities greatly outnumber those in our cluster field because the control field images are deeper. 
These tendencies are also observed in our previous high-$z$ cluster WL studies (e.g., \citealt{Jee2011}). 
Therefore, in the central region where both F606W and F814W are available, we decide to proceed with our WL study without any blue cluster contamination correction.

In the peripheral region, we only apply the magnitude cut ($25.0 < \mbox{F606W} < 28.0$) and the resulting distribution is shown in light gray. Obviously, this magnitude-only selection includes red galaxies that would have been discarded if F814W had been available.
The excess due to the inclusion of these red sources is clearly seen (dark gray vs. light gray in the top panel of Figure~\ref{fig:mag_distribution}). 
However, one should not incorrectly attribute the difference exclusively to the red cluster members because many non-cluster members can also have their color satisfying the condition $\mbox{F606W}-\mbox{F814W}> 1.0$. 
In \S\ref{section_beta}, we find that in fact the second issue dominates the excess and the red cluster member contamination is not significant. This is not surprising because the fraction of the early-type galaxies is expected to be low in the faint magnitude regime and also decreases rapidly with clustocentric radius. \\

{\bf Source selection with WFC3 NIR photometry}
Since all objects observed with WFC3 are located inside the central region and have color, we select the sources using both color and magnitude criteria: $\mbox{F105W}-\mbox{F140W}< 0.5$ and $\mbox{F140W}>23.5$. Similarly to the ACS-based selection, these thresholds in color and magnitude are chosen to select sources bluer and fainter than the locus of the red-sequence (bottom panel of Figure~\ref{fig:cmd}).
Again, we check for the presence of a possible contamination by blue cluster members by comparing the resulting magnitude distributions with control field statistics. The bottom panel of Figure~\ref{fig:mag_distribution} indicates that the source number density is consistent with that from HUDF at $\mbox{F140W}\lesssim26$. At $\mbox{F140W}\gtrsim26$, the source density becomes gradually lower because of the shallower depth.

\subsection{Source Redshift Estimation} \label{section_beta}
We determine the effective redshift and the width of the redshift distribution of our source population using the HUDF photo-$z$ catalog \citep{UVUDF}. 
Given that the field size of HUDF is small ($\mytilde12.8$~\sqarcmin), one might be concerned about the sample variance. 
We will show in \S\ref{sample_variance} that the mass uncertainty due to the sample variance (up to $\mytilde7$\% in mass) is much less than the statistical error.
As mentioned in \S\ref{phot_requirements}, we apply the same selection criteria to the HUDF galaxies. 

Because of the absence of ACS/F814W in the HUDF, 
we perform a photometric transformation to estimate the F814W magnitude using the HUDF F606W and F775W photometry. 
The transformation to F814W from F606W and F775W depends on the spectral energy distribution (SED) and redshift. 
The HUDF catalog includes photometry (AB magnitudes of ACS and WFC3 instruments), photometric redshift, and best-fit spectral type. 
Utilizing the information on redshift and best-fit spectral type (model SED), we perform synthetic photometry (multiplying the filter throughput to the redshifted SED) to obtain a mathematical relation among F814W, F606W, and F775W. 

We compute the lensing efficiency $\beta$ as follows:
\begin{equation}
\beta = \left < \mbox{max} \left (0,\frac{D_{ls}}{D_s} \right ) \right >
\end{equation}
\noindent
Note that for galaxies with redshift less than the cluster redshift, we assign zero because they are not lensed. 
The upper panel of Figure~\ref{fig:mean_beta} shows $\beta$ as a function of the F606W magnitude. As observed, $\beta$ should increase with magnitude because sources at higher redshifts are fainter on average. 
The blue filled circles represent the cases where we have color measurements (thus the red-sequence is removed). 
In the peripheral region, similar $\beta$ values are observed when only magnitudes are used for selection (red circles). 
When computing the representative value of $\left < \beta \right >$ for the entire source population, we apply proper weights to different magnitude bins because of the difference in depth between the HUDF and our cluster field using the following equation:
\begin{equation}
\left<\beta \right> = \frac{\mathop{\sum}\limits_{i}\beta_{i} w_{i} } {\mathop{\sum}\limits_{i}w_{i} }, \label{eqn_beta_estimation}
\end{equation}
\noindent
where $\beta_i$ is the average lensing efficiency at the i$^{th}$ magnitude bin in Figure~\ref{fig:mean_beta} and
$w_i$ is the weight needed to take into account the difference in depth, which is the ratio of the number density of our sources to that of the HUDF galaxies selected with the same criteria. 

In the peripheral region where a color is not available, we apply an additional correction for inclusion of red galaxies as follows.
This magnitude-based selection contains red background, red foreground, and red cluster galaxies as well as blue sources that would have been selected as the source population if the color had been available. These blue sources consist of blue foreground/background galaxies and blue cluster members. 
Since the red foreground and red cluster galaxies are not lensed, the lensing efficiency at each magnitude bin in the peripheral region is computed as below: 
\begin{equation}
\beta_{p,i} = \frac{n_{bs,i}} {n_{p,i}}\beta_{bs,i} + \frac{n_{rb,i}} {n_{p,i}}\beta_{rb,i} , \label{eqn_beta_peripheral}
\end{equation}
\noindent
where $\beta_{bs,i}$ ($n_{bs,i}$) and $\beta_{rb,i}$ ($n_{rb,i}$) are the average lensing efficiencies (number densities) at the i$^{th}$ magnitude bin for blue sources and red background galaxies, respectively. 
$n_{p,i}$ is the total number density within each magnitude bin in the peripheral region. 
$\beta_{bs,i}$ is the same as the $\beta_i$ value measured from the central region. 
$n_{bs,i}/n_{p,i}$ corresponds to the ratio of the number density in the central region (dark gray) to that in the peripheral region (light gray) of the cluster field in Figure~\ref{fig:mag_distribution}. 
The number density of red background galaxies in the peripheral region ($n_{rb,i}$ in the second term of Equation~\ref{eqn_beta_peripheral}) is not directly observable. 
Nevertheless, we can approximate the ratio $n_{rb,i}/n_{p,i}$ using the HUDF photo-$z$ catalog. Since the catalog does not contain red cluster members, in principle this approximation leads to overestimation. However, the fraction of red cluster members is estimated to be small ($< 5$\%) in the peripheral region. Thus, the overestimation is negligibly small ($\mytilde4.7$\%).

Using the above methods, we obtain $\left < \beta \right >=0.201$ for the central region corresponding to an effective source plane redshift $z_{\rm eff}=1.591$, which is translated to the critical surface density $\Sigma_{c}\simeq4752~M_{\sun}\mbox{pc}^{-2}$.
For the peripheral region (magnitude selection only), $\left < \beta \right >$ is $0.139$ corresponding to $z_{\rm eff}=1.413$ ($\Sigma_{c}\simeq6865~M_{\sun}\mbox{pc}^{-2}$). If we had ignored the correction due to red source inclusion, we would have obtained a $\mytilde40$\% higher value of $\left < \beta \right >=0.196$. 

Although this single source plane assumption is convenient, ignoring the distribution width creates non-negligible bias in our cluster mass estimation because the lensing efficiency is a nonlinear function of source redshift. 
We obtain the width of the redshift distribution to be $\left < \beta^{2} \right >=0.070$ and $0.048$ for the central and peripheral region, respectively. 
\cite{Seitz1997} derived the first-order correction to be:
\begin{equation}
\frac{g\prime}{g} = 1 + \left
(\frac{\left < \beta^{2} \right >}{\left<\beta\right>^{2}} - 1 \right )\kappa
\end{equation}
\noindent
where $g\prime$ and $g$ are the observed and true shears, respectively.

For the sources detected in F140W, we follow the same procedure using the HUDF photo-$z$ catalog. 
Similarly to the case for F606W, the mean $\beta$ value increases with magnitude as shown in the lower panel of Figure~\ref{fig:mean_beta}. 
We obtain $\left< \beta \right>=0.161$, $\left < \beta^{2} \right >=0.057$, and $z_{\rm eff}=1.470$ for the F140W sources. \\ \\

\begin{figure*}
\centering
\includegraphics[width=180mm, height=180mm]{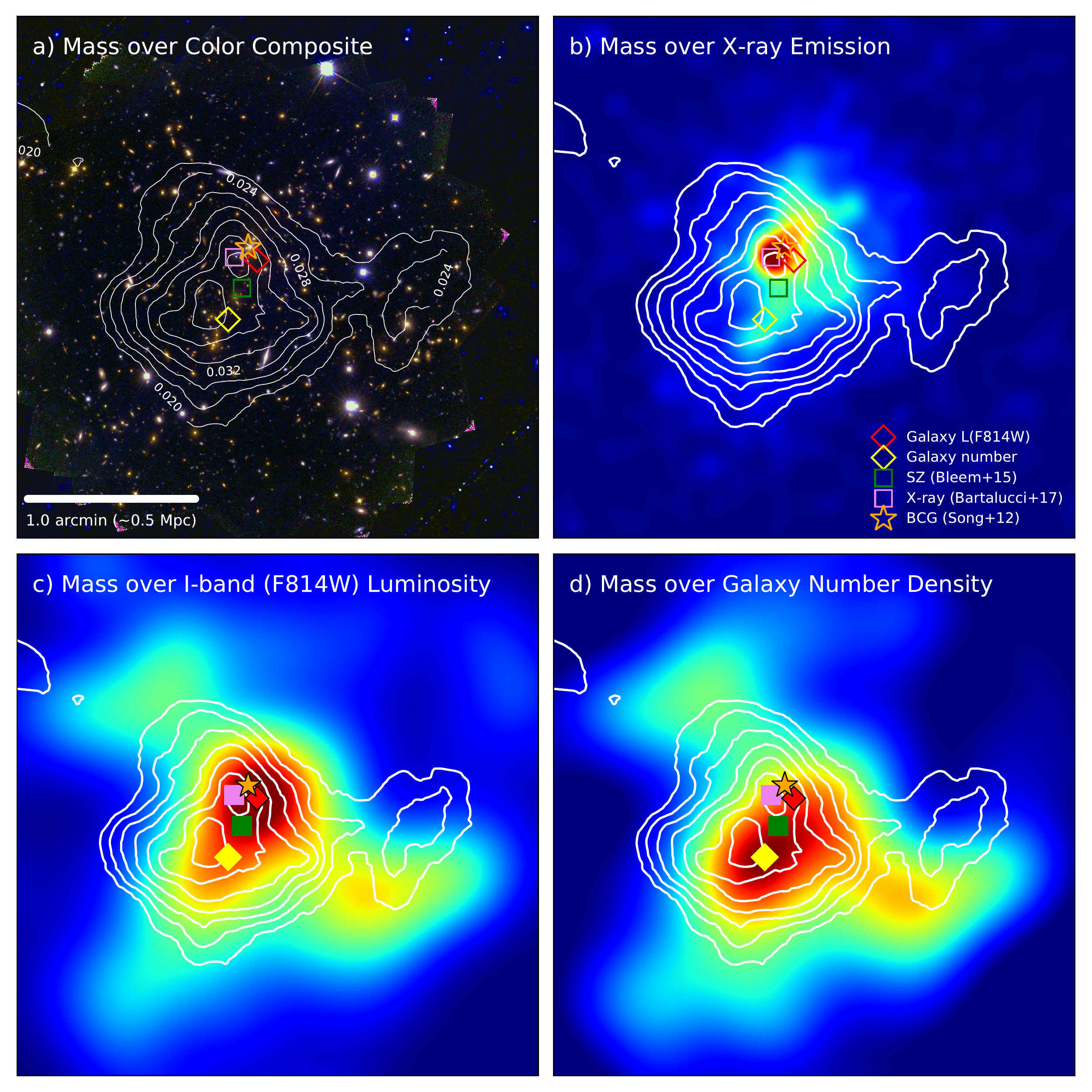}
\caption{
Two-dimensional mass reconstruction of SPT2106 using the {\tt FIATMAP} code \citep{FIATMAP}. 
All figures show the central $3\arcmin\times3\arcmin$ region.  
The symbols represent different centroids that we use to investigate the impact of this choice on mass measurement (see text). 
a) Mass contours overlaid on the color composite. This color composite is created with \HST~ACS/WFC F606W (blue), ACS/WFC F814W (green), and WFC3/IR F105W (red). The outermost contour corresponds to the $2.5\sigma$ significance and the significance levels increase inward by $0.5\sigma$. The contour label is convergence $\kappa$. Since we do not break the mass-sheet degeneracy, the scale is somewhat arbitrary. 
The mass distribution consists of two mass clumps: ``main" and ``western" extension. 
The main clump is also resolved into the northwestern and southeastern subclumps. 
The southeastern subclump has a slightly larger significance than the northwestern one. 
b) Same as a) except that the mass contours are overlaid on the {\it Chandra} X-ray image 
processed by CIAO package with removing point sources. 
The X-ray emission is highly elongated and the peak coincides with the northwestern subclump.
c) Same as b) except that the mass contours are overlaid on the cluster galaxy candidate F814W luminosity density.
The smoothing scale of the luminosity density map is FWHM~$\sim37\arcsec$. 
d) Same as a) except that the mass contours are overlaid on the number density map (smoothed with the same kernel applied to the luminosity density map). 
}
\label{fig:DM+baryons}
\end{figure*}

\begin{figure}
\centering
\includegraphics[width=80mm, height=120mm]{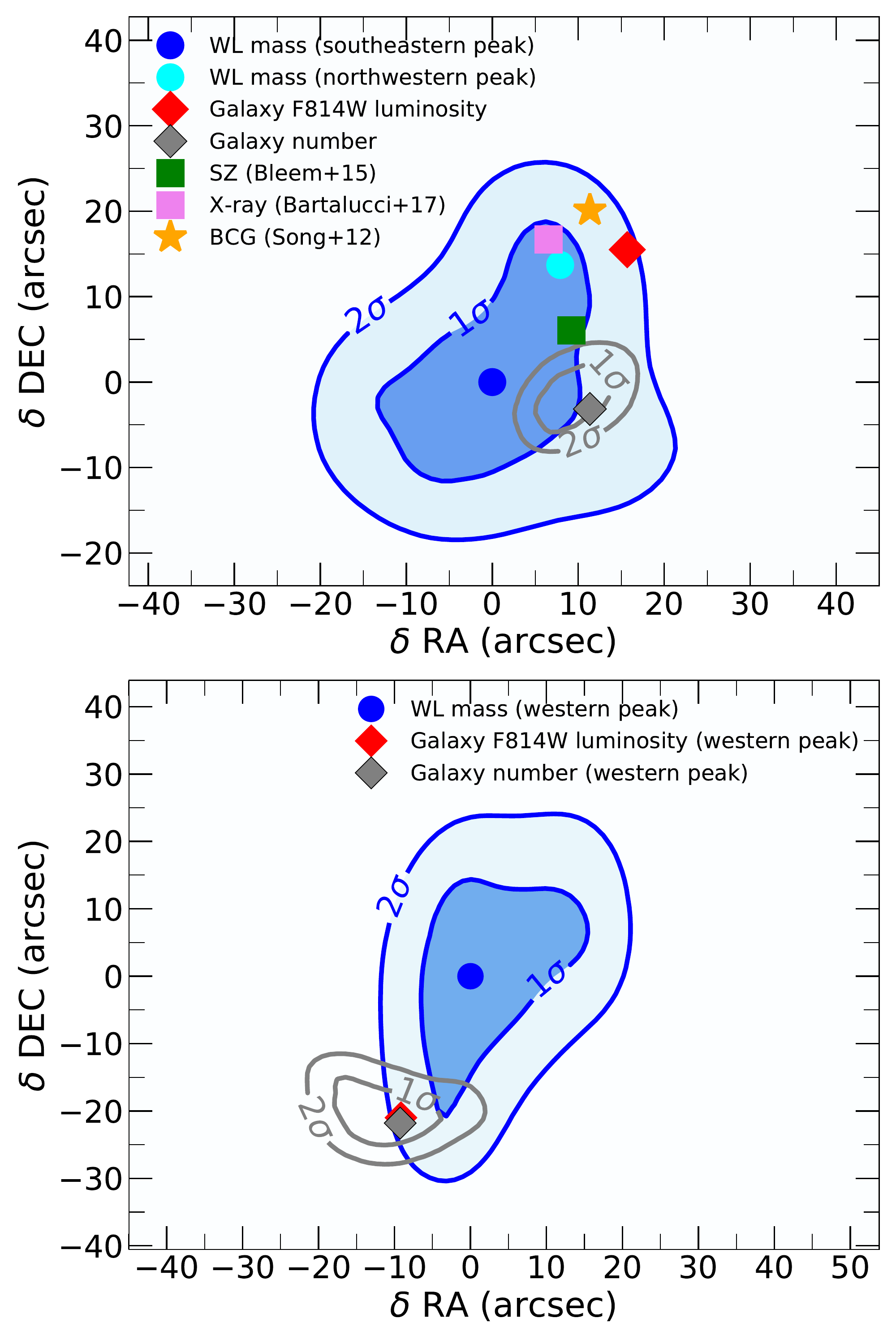}
\caption{WL mass centroid significance test
for the main (top) and western (bottom) clumps. The blue contours represent the centroid distributions for the main (top) and western (bottom) mass clumps in our 1000 bootstrapping runs. 
The X-ray, SZ, galaxy luminosity, and galaxy number density peaks are located near the $1\sigma$ centroid contour of the main clump. 
We also display the centroid distributions of the number density for both the main and western clumps with gray contours from our bootstrapping resamples. 
The number density centroids are statistically consistent with WL mass centroids. 
}
\label{fig:centroids}
\end{figure}

\section{Results} \label{section_results}

\subsection{Mass Reconstruction} \label{mass_map}
We reconstruct the mass distribution of the cluster by averaging ellipticities of the background sources and then convolving the ellipticity map using the relation below: 
\begin{equation}
\kappa (\bvec{x}) = \frac{1}{\pi} \int d^2 \bvec{x}~D^*(\bvec{x}-\bvec{x}^\prime) \gamma (\bvec{x}^\prime) \label{k_of_gamma},
\end{equation}
where $D^{*}(\bvec{x})$ is the complex conjugate of the convolution kernel defined as $D(\bvec{x} ) = - 1/ (x_1 - i x_2 )^2$ and $\gamma (\bvec{x})$ is the mean ellipticity (shear) distribution of the background sources. For our mass reconstruction,
we use the {\tt FIATMAP} code \citep{FIATMAP}, which implements the convolution in real space.
We verify that this {\tt FIATMAP} result is highly consistent with the one derived from our implementation of the \cite{KS93} algorithm in Fourier space.
Since the reconstruction is a noisy process, it is important to assess the significance of the observed features.
For this significance assessment, we generate an uncertainty (rms) map of the mass reconstruction by
performing 1000 bootstrap resamples with randomly (while allowing redundancy) selected source galaxies.

Figure~\ref{fig:DM+baryons} shows our mass reconstruction. We create overlays with the color-composite, X-ray emission, F814W luminosity, and number density distributions. 
The luminosity and number density maps are made by selecting cluster member candidates,
which meet the following criteria: $24 < \mbox{F606W} < 26$ and $1.4 <\mbox{F606W}-\mbox{F814W}< 2.2$ (see also Figure~\ref{fig:cmd}); the brightest cluster galaxy \citep[BCG;][]{Song2012} at (RA,DEC)=(21:06:04.66,-58:44:28.32) is manually included (not initially selected by the criteria because of its extreme luminosity F606W~$\sim22$). 

The X-ray emission map is obtained from the {\it Chandra} archive\footnote{http://cxc.harvard.edu/cda/}. 
The observation was performed in 2010 and 2011 (ObsID of 12180 and 12189, respectively) with {\tt VFAINT} mode using the ACIS-I instrument and the total exposure time of $\mytilde73$ ks. 
The {\it Chandra} X-ray data reduction is conducted using the {\it Chandra} Interactive Analysis of Observations (CIAO; \citealt{CIAO}) ver. 4.11.
After combining and reprojecting to the same tangent plane using the \texttt{merge$\_$obs} tool, 
we create an X-ray emission map within the energy range of $[0.5-7.0]$ keV and obtain an exposure-corrected map. 
We use the \texttt{wavdetect} and \texttt{aconvolve} tools to remove point sources and to adaptively smooth the map, respectively. The morphological characteristics in the WL mass reconstruction, the cluster galaxies and the X-ray emission are discussed below.

\subsubsection{Comparison with Cluster Galaxies} \label{comp_members}
Our high-resolution mass reconstruction enables a detailed comparison with the cluster galaxy distributions. 
Both luminosity and number density maps show that the galaxy distribution is not symmetric and can be described as consisting of the main clump and the western extension. 
Our WL mass map clearly detects these two structures. 
As observed in both luminosity and number density maps, 
the mass clump corresponding to the western galaxy concentration is also much weaker than the main mass clump. Our bootstrapping analysis shows that the significance of the western mass clump is $\mytilde3.4\sigma$. 

In Figure~\ref{fig:DM+baryons}, the distribution of the western galaxies (both luminosity and number density) is offset from that of the mass contours by $\mytilde30$\arcsec~(distance between the two centroids). 
We investigate the centroid uncertainties of mass and galaxies in the western substructure by using bootstrapping analysis. 
The centroid is determined by two-dimensional circular Gaussian fitting. Among the parameters that we described in \S\ref{section_shape_measurement}, only centroid (x,y) is set free; 
we set zero for background, the highest pixel value for amplitude, and the effective smoothing scales for the semi-major and -minor axes. 
The bottom panel of Figure~\ref{fig:centroids} shows that the $1\sigma$ contours for mass and galaxy touch each other. 

The mass map further resolves two substructures within the main mass clump. Our bootstrapping analysis shows that the northwestern (southeastern) peak is detected with a significance of $\mytilde5.7\sigma$ ($\mytilde6\sigma$).
The orientation of the vector connecting the two substructures is similar to the direction (northwest to southeast) of the elongation seen in the galaxy distributions. The distance between the two substructure centroids is $\mytilde150$~kpc ($\mytilde20$\arcsec). The effective smoothing scale of our mass reconstruction is FWHM~$\sim17$\arcsec. If we increase the effective smoothing scale above $\mytilde20$\arcsec, the two peaks merge into a single peak elongated in the northwest-southeast direction. 

Interestingly, the centroid of the northwestern substructure lies near the luminosity peak whereas that of the southeastern substructure is closer to the number density peak. The difference between the luminosity and number density distributions arises from the BCG being much more luminous than the other member candidates by at least $\mytilde2$ magnitude, although more galaxies are concentrated near the southeastern WL peak.
Figure~\ref{fig:centroids} shows our significance test of the mass peak centroids based on bootstrapping.
An important caveat to remember in comparison with cluster galaxies is that our galaxy map is based on photometric selection targeted for the red sequence and subject to non-member contamination and omission of blue members. 
Currently, the number of the spectroscopically confirmed members is 18 \citep{Foley2011}.

\subsubsection{Comparison with X-ray Emission} \label{comp_Xrays} 
The asymmetric X-ray surface brightness distribution of SPT2106 was considered as an indication of merger activity in F11. 
Comparison of the X-ray image with our mass reconstruction and galaxy distributions also supports this merger hypothesis.
As seen in Figure~\ref{fig:DM+baryons}b, the {\it Chandra} image shows an elongated X-ray emission whose peak location is consistent with the northwestern mass substructure centroid. No strong X-ray emission is found in the other substructure. 
In some cluster mergers with small impact parameters, a bright cool core is often found near the less massive system while there is relatively very weak X-ray emission around the more massive (thus hot) component 
(e.g., \citealt{Jee2014}; \citealt{Golovich2017}; \citealt{vanWeeren2017}).
Also, the elongation of the X-ray emission is consistent with the orientation of the vector connecting the two substructures seen in both mass and galaxy distributions. Thus, the absence of any significant X-ray peak near the southeastern substructure may indicate that the two substructures might have passed through each other.

\begin{figure*}
\centering
\includegraphics[width=180mm]{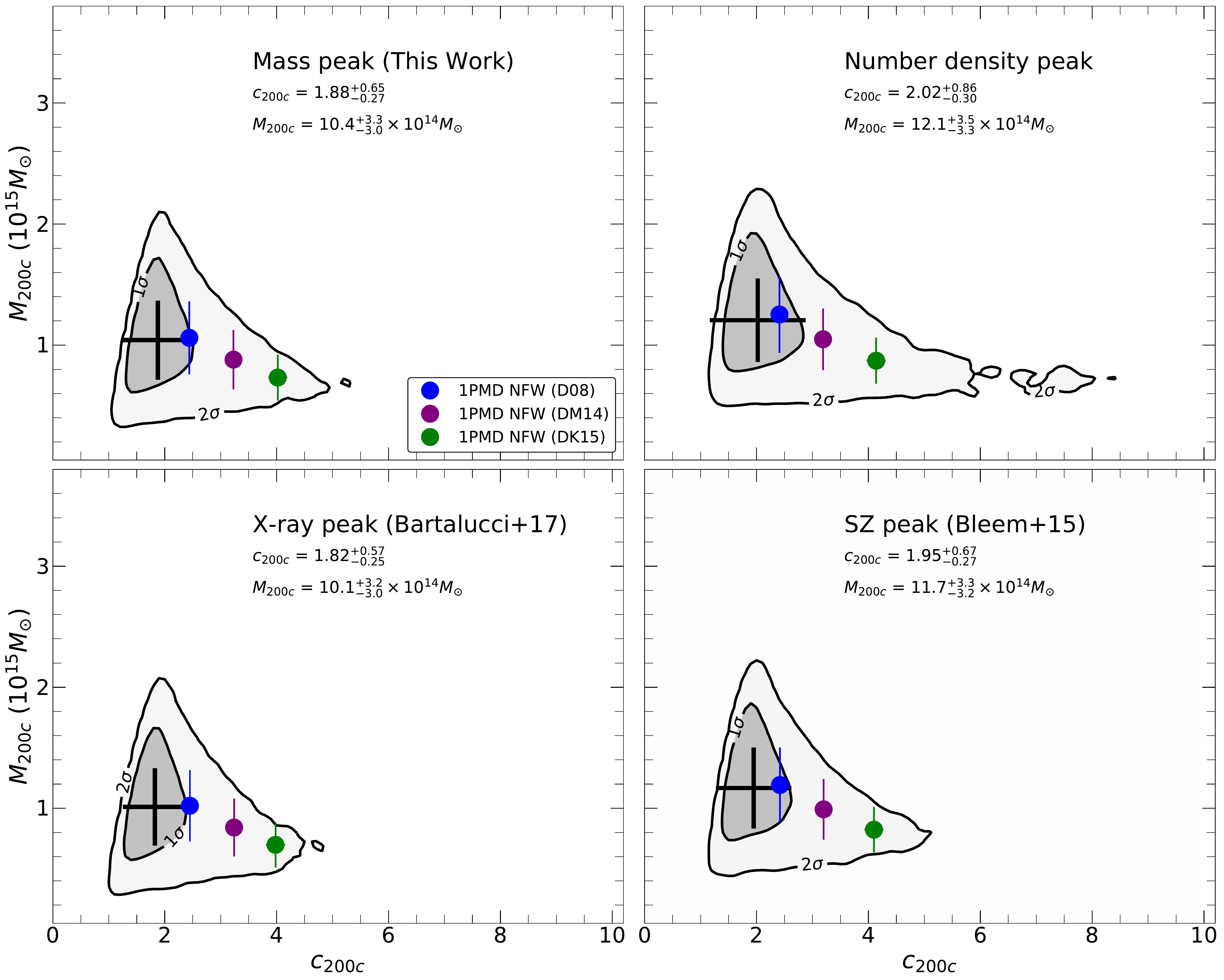}
\caption{
Posterior distribution from 500,000 MCMC samples of our two-parameter mass determination.
We consider the four centroids: WL mass, number distribution of the cluster member candidates, X-ray, and SZ peak. 
The best-fit values and marginalized uncertainties of concentration $c_{200c}$ and mass $M_{200c}$ are indicated by black crosses.
We also compare the results obtained from three mass-concentration relations using one-parameter fitting (1PMD, color circles). 
}
\label{fig:MCMC_figs}
\end{figure*}

\begin{figure*}
\centering
\includegraphics[width=180mm]{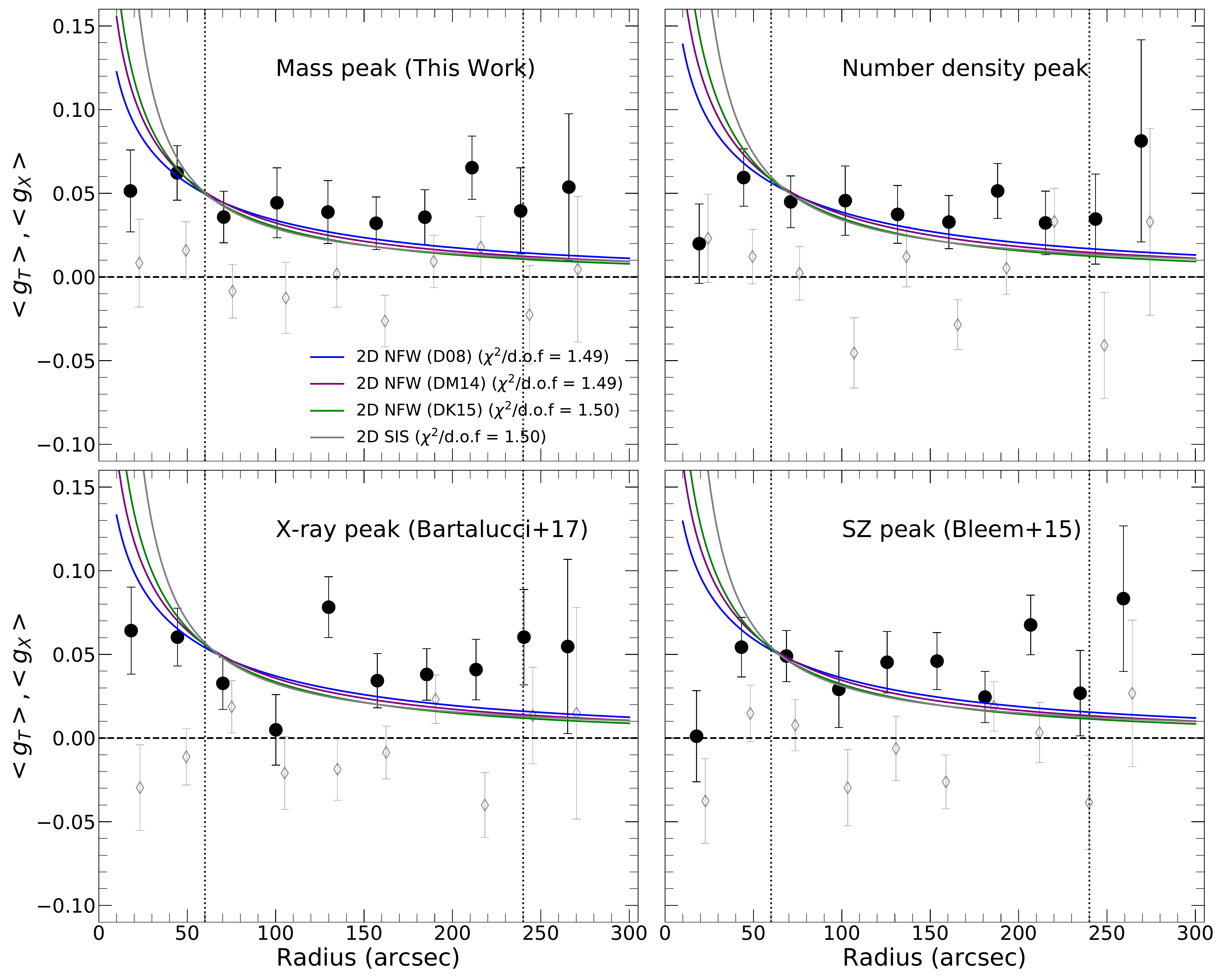}
\caption{
Reduced tangential shear profiles centered at four different centroids. 
Filled circles represent the tangential shear ($g_{T}$) while open diamonds show the cross shear ($g_{X}$, the 45\degr~rotation of the source images).
The vertical dotted lines are the cut-off radii. 
We show the best-fit models using the singular isothermal sphere (SIS, gray) and NFW halos with three different mass-concentration relations; we use blue, purple, and green for the \cite{Duffy08}, \cite{DM14}, and \cite{DK15} relations, respectively. 
}
\label{fig:shear_profiles}
\end{figure*}

\subsection{Mass Estimation} \label{mass_estimate}
We derive the cluster mass by assuming that the radial shear profile is approximated by the Navarro-Frenk-White (NFW, \citealt{NFW1997}) model. The NFW profile describes the average halo distribution in dark matter numerical simulations and has been a popular choice in many WL mass studies. However, if our cluster deviates substantially from the profile, this will give rise to a model bias. We will discuss the impact of this model bias and other miscellaneous sources on our mass uncertainty in \S\ref{systematic_errors}. 
Although an on-going merger activity is indicated from our WL mass map, cluster member candidates, and X-ray emission, we proceed with a single halo approximation because the two substructures in the main clump are too close to be modeled as separate halos ($\mytilde150$kpc). 
We will also deal with its impact on the mass systematics in \S\ref{systematic_errors}.

Since the two free parameters (e.g., concentration $c$ and scale radius $r_s$) of the NFW profile are highly correlated with each other, 
it is difficult to determine the two parameters simultaneously with noisy data. Therefore, many authors
perform one-parameter fitting assuming the two parameters follow a tight relation. Typically, the relation is expressed
in terms of the evolution of the correlation between cluster mass and concentration with redshift.
Since this so-called mass-concentration ($M-c$) relation is only the average behavior and individual clusters show large scatter around the mean relation,
using the relation is an additional source of systematics in cluster mass estimation.
In this paper, although we also present results from this traditional method to enable comparison with previous studies, our main result is obtained
without assuming any particular $M-c$ relation. This requires us to perform a Markov Chain Monte Carlo (MCMC) analysis with the two
parameters: scale radius $r_s$ and concentration $c$,
and compute a marginalized cluster mass (we refer to this two-parameter fitting result as 2PMD hereafter). The resulting statistical uncertainty is larger than when one uses the $M-c$ relation (hereafter 1PMD).

Typical WL studies first construct a one-dimensional (1D) radial tangential shear profile and then fit a model to it. This makes the result sensitive to the binning choice. To avoid this binning effect, we perform a two-dimensional (2D) model fitting to individual galaxy shapes without applying any binning scheme. Our log-likelihood function $\mathcal{L}$ is given as \citep{Mincheol2019}:
\begin{equation}
\mathcal{L} = \sum_{i} \sum_{s=1,2} \frac{ [ g^m_s (M_{SPT2106},x_i,y_i,z_s) - g^o_s(x_i,y_i)]^2 } {\sigma_{SN}^2 + \sigma_e^2},
\end{equation}
where $g^m_s$ ($g^o_s$) is the $s^{th}$ component of the predicted (observed) reduced shear for the source redshift $z$ at the $i^{th}$ galaxy position $(x_i,y_i)$ as a function of the cluster mass $M_{SPT2106}$. 
The ellipticity dispersion (shape noise) is $\sigma_{SN}=0.25$ whereas $\sigma_e$ is the ellipticity measurement noise of each object. 
In addition to the binning effect removal, this method allows us to assign individual redshifts to source galaxies. This merit is important
in the current study because we combine source catalogs from different selection schemes, which thus produce three different effective redshifts. 

For each source, the tangential shear is obtained as below:
\begin{equation}
g_T = -e_1 \cos 2\theta - e_2 \sin 2\theta \label{tan_shear},
\end{equation}
\noindent
where $e_{1(2)}$ is the ellipticity component 
and $\theta$ is the position angle of the source with respect to the reference cluster center.
Obviously, the exact shape of the profile depends on the choice of the reference cluster center.
We use the following four reference centers: the global peak location of our WL mass (the southeastern subclump), the number density center of the early-type member candidates, the X-ray emission peak \citep{Bartalucci2017}, and the SZ decrement centroid \citep{Bleem2015}.
The coordinates of these positions are listed in Table~\ref{table_2PMD_MCMC} and indicated with various symbols in Figure~\ref{fig:DM+baryons}.

We limit the fitting range to $60\arcsec<r<240\arcsec$. 
The lower limit is needed because of a number of issues (e.g., \citealt{Jee2011}). Two of the most important reasons are 1) that the cluster member contamination may be most significant very near the cluster center and 2) that the location of the true cluster center is unknown. 
Outside the upper limit ($r > 240$\arcsec), the annulus cannot complete a circle due to the limits of our imaging coverage.

\subsubsection{Two-parameter MCMC analysis} \label{mass_estimate_2PMD}
We use the open-source python package {\tt emcee} \citep{emcee}
when performing our MCMC. 
We use flat priors for both scale radius $r_s$ and concentration $c$.
The prior interval of $c$ is set to $0 < c < 10$. 
We verify that the upper limit is sufficiently high and most of our MCMC samples stay below this upper limit.
The lower limit is set to zero 
because the concentration in the NFW halo, by definition, should have a positive value. 
We limit the range of the scale radius to $10\arcsec < r_s < 110\arcsec$. 
Together with the interval of the concentration parameter, this $r_s$ range results in cluster masses bound within $\mytilde10^{11}~M_{\sun}$ to $\mytilde10^{17}~M_{\sun}$.

Figure~\ref{fig:MCMC_figs} shows the posterior distributions from our 500,000 MCMC samples in $M_{200c}$ and $c_{200c}$ for the four choices of the cluster center.
All four results are consistent and show that SPT2106 is indeed an extremely massive system.
For example, when we select our WL mass peak as the cluster center, we obtain $M_{200c} = 10.4^{+3.3}_{-3.0}~\times$~\solarm. 
The marginalized masses are summarized in Table~\ref{table_2PMD_MCMC}.

\begin{table*}
\begin{center}
\caption{Impacts of the centroid choice on the mass estimation in our MCMC analysis \label{table_2PMD_MCMC}}
\begin{tabular}{cccccccc}
\tableline
\tableline
\\
 \colhead{Centroids} & \colhead{R.A.} & \colhead{DEC.} & \colhead{Concentration} & \colhead{$R_{200c}$} & \colhead{$M_{200c}$} & \colhead{$R_{500c}$} & \colhead{$M_{500c}$} \\ 
 & \colhead{(J2000)} & \colhead{(J2000)} & & \colhead{(Mpc)} & \colhead{($10^{14}~M_{\sun}$)} & \colhead{(Mpc)} & \colhead{($10^{14}~M_{\sun}$)} \\ [0.1ex]
\hline
\\[0.1ex]
WL mass$^1$ & 21:06:06.44 & -58:44:46.36 & $1.88^{+0.65}_{-0.27}$ & $1.4^{+0.1}_{-0.1}$ & $10.4^{+3.3}_{-3.0}$ & $0.8^{+0.1}_{-0.1}$ & $6.0^{+1.9}_{-1.7}$ \\ [1.2ex] 
Galaxy Number$^2$ & 21:06:05.56 & -58:44:53.16 & $2.02^{+0.86}_{-0.30}$ & $1.5^{+0.1}_{-0.1}$ & $12.1^{+3.5}_{-3.3}$ & $0.9^{+0.1}_{-0.1}$ & $7.1^{+1.9}_{-1.8}$ \\ [1.2ex] 
X-ray$^3$ & 21:06:05.28 & -58:44:31.70 & $1.82^{+0.57}_{-0.25}$ & $1.4^{+0.1}_{-0.2}$ & $10.1^{+3.2}_{-3.0}$ & $0.8^{+0.1}_{-0.1}$ & $5.7^{+1.9}_{-1.7}$ \\ [1.2ex] 
SZ$^4$ & 21:06:04.94 & -58:44:42.36 & $1.95^{+0.67}_{-0.27}$ & $1.4^{+0.1}_{-0.1}$ & $11.7^{+3.3}_{-3.2}$ & $0.9^{+0.1}_{-0.1}$ & $6.8^{+1.9}_{-1.7}$ \\ [1.2ex]
\hline
\hline
\tableline
\end{tabular}
\end{center}
\tablecomments{1. WL mass center (the southeastern peak). 2. Galaxy number density peak. 3. X-ray emission peak \citep{Bartalucci2017}. 4. SZ centroid \citep{Bleem2015}.}

\end{table*}

\subsubsection{One-parameter tangential shear fitting} \label{mass_estimate_1PMD}
Although we claim that our two-parameter result presented in \S\ref{mass_estimate_2PMD} is the least biased mass, here we present
our mass estimates based on one-parameter fitting method to enable fair comparisons with previous studies. 
We consider the three mass-concentration relations from \cite{Duffy08} (hereafter D08), \cite{DM14} (hereafter D14), and \cite{DK15} (hereafter D15). 
In Figure~\ref{fig:shear_profiles} we display the radial tangential shear profile for the four different centers. 
The resulting masses are summarized in Table~\ref{table_1PMD} for the three $M-c$ relations and also for the singular isothermal sphere (SIS) model.
Our results show that the cluster mass is insensitive to the choice among the four centers considered in this paper whereas the profile assumption is an important systematic factor. 
We obtain the highest masses for the D08 relation and the lowest masses for the DK15 relation, although their error bars marginally overlap.

\begin{table*}
\begin{center}
\caption{Mass estimates based on one-parameter 2D fitting for various choices of centroids and mass-concentration relations. \label{table_1PMD}}
\begin{tabular}{cccccc}
\tableline
\tableline
       & \multicolumn{2}{c}{2D SIS} & \multicolumn{3}{c}{2D NFW} \\ 
 \colhead{Centroids} & \colhead{$\sigma_{v}$} & \colhead{$M_{200c,SIS}$} & \colhead{$M_{200c,D08}$} & \colhead{$M_{200c,DM14}$} & \colhead{$M_{200c,DK15}$} \\ 
 & \colhead{(\kms)} & \colhead{($10^{14}~M_{\sun}$)} & \colhead{($10^{14}~M_{\sun}$)} & \colhead{($10^{14}~M_{\sun}$)} & \colhead{($10^{14}~M_{\sun}$)} \\ [0.1ex]
\hline
\\[0.1ex]

WL mass & $1152\pm104$ & $7.7^{+2.3}_{-1.9}$ & $10.6^{+3.5}_{-2.6}$ & $8.8^{+2.8}_{-2.1}$ & $7.3^{+1.9}_{-1.9}$ \\ [1.2ex]
Galaxy Number & $1227\pm93$ & $9.3^{+2.3}_{-2.0}$ & $12.5^{+3.6}_{-2.7}$ & $10.5^{+2.9}_{-2.2}$ & $8.7^{+1.9}_{-1.9}$ \\ [1.2ex]
X-ray & $1134\pm105$ & $7.4^{+2.3}_{-1.9}$ & $10.2^{+3.4}_{-2.5}$ & $8.4^{+2.8}_{-2.0}$ & $7.0^{+1.9}_{-1.9}$ \\ [1.2ex]
SZ & $1202\pm95$ & $8.8^{+2.3}_{-1.9}$ & $11.9^{+3.5}_{-2.7}$ & $9.9^{+2.8}_{-2.2}$ & $8.2^{+1.9}_{-1.9}$ \\ [1.2ex]

\hline
\hline
\tableline
\end{tabular}
\end{center}
\end{table*}

\section{Discussion} \label{section_discussion}

\subsection{Comparison with other non-WL studies} \label{mass_comparison}
Since its discovery from the initial SPT survey, there have been several efforts to improve the mass measurement of SPT2106. 
The mass constrained by SZ data (e.g., \citealt{Williamson2011}; \citealt{Reichardt2013}; \citealt{Bleem2015}) ranges from $M_{200c}=9.0$~to~$10.6~\times$~\solarm~($M_{500c}=5.4$ to $8.4~\times$~\solarm). 
The broad range of these mass estimates 
stems from different selections of SZ data (single band vs. multiple bands) and/or different scaling relations.
Studies with X-ray scaling relations (F11; \citealt{Ruel2014}; \citealt{Bartalucci2017}) suggest somewhat higher masses ($M_{200c}\sim18~\times$~\solarm), although the estimate by \cite{Amodeo2016} based on the hot gas temperature profile using the NFW model gives a result nearly a factor of two smaller. 
Based on the velocity dispersion, F11 measure the mass of the cluster to be $M_{200c}=1.4^{+1.7}_{-0.8}~\times~10^{15}~M_{\sun}$. 
Although this mass is consistent with our WL estimate, 
we believe that both the small number (18 members) and the possible merger activity described in \S\ref{mass_map} make it difficult to infer the cluster mass based on the velocity dispersion alone. 

\begin{figure}
\centering
\includegraphics[width=80mm, height=80mm]{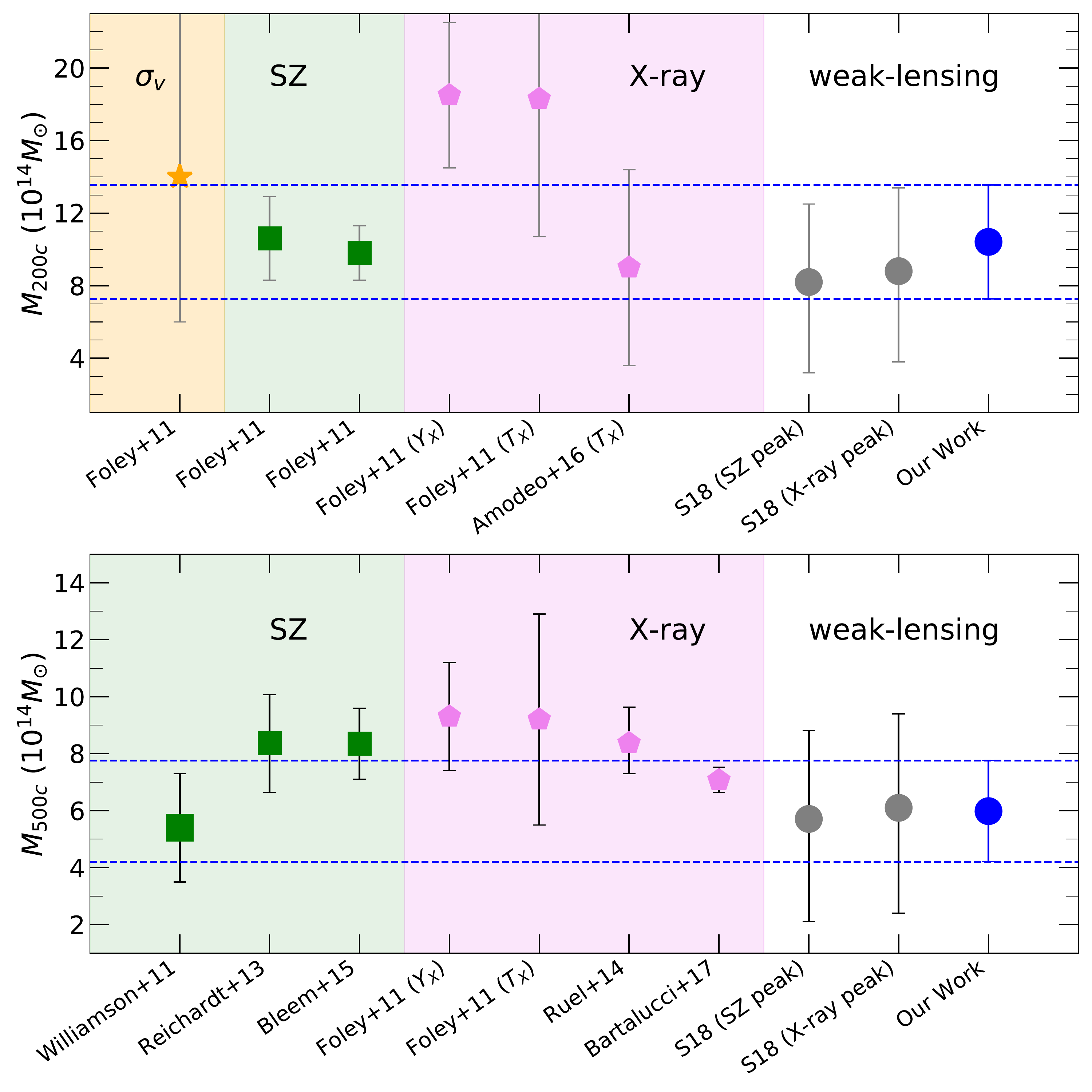}
\caption{
Mass comparison (Top: $M_{200c}$, Bottom: $M_{500c}$) of SPT2106~measured with various proxies. 
We choose the two-parameter MCMC result measured at our WL mass peak to represent our work. 
Our WL mass estimate is consistent with previous measurements. 
}
\label{fig:mass_compare}
\end{figure}

In Figure~\ref{fig:mass_compare} we present a summary of these previous results (for both $M_{200c}$ and $M_{500c}$) in chronological order. 
Also displayed is the previous WL measurement by S18, whose central value is somewhat lower than our value, although the statistical error bars overlap. Detailed discussions on the comparison of the WL results are presented in \S\ref{mass_comparison_S18}. 
We find that our WL mass is consistent with the previous results.

\subsection{Comparison with the S18 WL result} \label{mass_comparison_S18}
SPT2106 was included in the sample of the 13 high-redshift clusters studied by S18 with WL. Their mass of SPT2106 is $8.8^{+5.0}_{-4.6}~\times$~\solarm, whose central value is $\mytilde85$\% of our result. Also, their mass uncertainty is $\mytilde50$\% larger than ours.
Although their error bars overlap with ours (our central value corresponds to the $1\sigma$ upper limit of the S18 result), the difference deserves scrutiny because S18 and our study share the same \HST/ACS imaging data. 
Therefore, in this section we provide detailed comparisons and attempt to trace the origin of the discrepancies. We focus on the following four aspects that can sensitively affect the cluster mass measurement and its uncertainty: 1) shear calibration, 2) source redshift estimation, 3) mass-concentration relation, and 4) source selection.

\subsubsection{Shear Calibration} \label{section_shear_calibration}
S18 use a moment-based shape measurement approach while our shape measurement is based on a model-fitting method. The moment-based approach was pioneered by \cite{KSB1995} (hereafter KSB) and has been modified by a number of authors (e.g., \citealt{Hoekstra1998}; \citealt{Erben2001}; \citealt{Schrabback2007}). 
The specific KSB branch used by S18 is described in detail by \cite{Schrabback2010}, where the authors apply the method to measure cosmic shears. 
One important difference made in S18 from the \cite{Schrabback2010} version is the use of the pixel-based CTE-degradation correction of \cite{Massey2014}. 
In \cite{Schrabback2010}, the correction is applied at the catalog level. 
The use of the pixel-based CTE-degradation is similar to our case, although our WL pipeline employs the STScI pipeline correction, which implements the \cite{Ubeda2012} algorithm. The fidelity test of this STScI correction in the context of WL is presented in \cite{Jee2014}.

Since the galaxy shape catalog of S18 is not public, it is impossible to compare shape measurements for individual galaxies. Nevertheless, it is possible to compare aggregate statistics when we closely mimic the S18 tangential shear measurement procedure. We match their source selection, source redshift estimation ($\left< \beta \right>=0.282$), shape measurement filter (ACS/F606W), applied cosmology ($\Omega_M=0.3$, $\Omega_{\Lambda}=0.7$, and $h=0.7$), reference point (X-ray peak from \citealt{Chiu2016}), and $M-c$ relation \citep{DK15} between the two analyses. We read off the S18 tangential shears from Figure~G8 of their paper. Because we compare the shears only in the region where both F606W and F814W colors are available, we choose the maximum distance to be $100$\arcsec. Although S18 exclude the shear at $\theta<60$\arcsec~in their mass estimation, we use all shear profiles at $\theta<100$\arcsec~to increase the statistical significance. This is justified because we are only interested in comparing the amplitude of the shears (shear calibration) between the two studies.
The resulting mass difference becomes $\mytilde10$\% (our mass is lower). This discrepancy is $\mytilde16$\% of the statistical error.

\subsubsection{Source redshift estimation.} \label{section_source_redshift}
S18 estimate the lensing efficiency $\left< \beta \right>$ using the photometric redshift catalog from the Cosmic Assembly Near-infrared Deep Extragalactic Legacy Survey (CANDELS, \citealt{CANDELS}). They apply the same source selection criteria to the CANDELS photo-$z$ catalog and obtain $\left< \beta \right>=0.282$ for their source population. 
When we apply the S18 color and magnitude selection to the CANDELS photo-$z$ catalog, we obtain $\left< \beta \right>=0.260$, which would increase the cluster mass by $\mytilde8$\%.
We believe that the discrepancy of $\left< \beta \right>\sim0.02$ is due to the difference in the shape criteria.
A larger difference ($\left< \beta \right>=0.323$) is found when we replace the CANDELS catalog with the HUDF one. This higher $\left< \beta \right>$ value would lower the cluster mass by $\mytilde15$\%.
The difference in $\left< \beta \right>$ between CANDELS and HUDF may be attributed to the systematics in the photo-z catalogs or the sample variance. 
Note that since we use the HUDF catalog for our analysis and our final mass (central value) is higher than the S18 one by $\mytilde18$\%, the mass shift due to the $\left< \beta \right>$
increase is in the opposite direction.

\subsubsection{Mass-concentration relation} 
S18 perform their one-dimensional tangential shear fitting utilizing only the DK15 $M-c$ relation, whereas we present results using three $M-c$ relations and also the one without the $M-c$ relation.
As shown in \S\ref{mass_estimate_1PMD} and Table~\ref{table_1PMD}, our mass estimates become lowest when the DK15 relation is used. The D08 and DM14 relations give larger mass estimates than the DK15 relation by $\mytilde21$\% and $\mytilde10$\%, respectively. We confirm the same trend also with the S18 tangential shears (read off from Figure G8 of their paper).
S18 reports $8.8^{+5.0}_{-4.6}~\times$~\solarm~using the DK15 relation and X-ray center. For the same choices, our mass estimate is $(7.0\pm1.9)~\times$~\solarm. Our central value is lower than the S18 value by $\mytilde20$\%. Since in \S\ref{section_shear_calibration} and \S\ref{section_source_redshift} we show that the differences in shear calibration and
$\left< \beta \right>$ estimation make our mass lower than the S18 one by $\mytilde10$\% and $\mytilde15$\%, respectively, it is plausible that this $\mytilde20$\% difference in mass might be attributed to the combined effect of the two.

As mentioned in \S\ref{mass_estimate_1PMD}, 
we regard the result obtained without any $M-c$ relation as the main mass estimate because we believe that it is the most unbiased value. 
When we use the X-ray center, the central value of this result ($10.1^{+3.2}_{-3.0}~\times$~\solarm) is $\mytilde45$\% higher than the one obtained from the DK15 relation [$(7.0\pm1.9)~\times$~\solarm]. Consequently, the difference in the $M-c$ relation is the leading cause of the discrepancy between ours and S18 results.

\begin{figure}
\centering
\includegraphics[width=80mm]{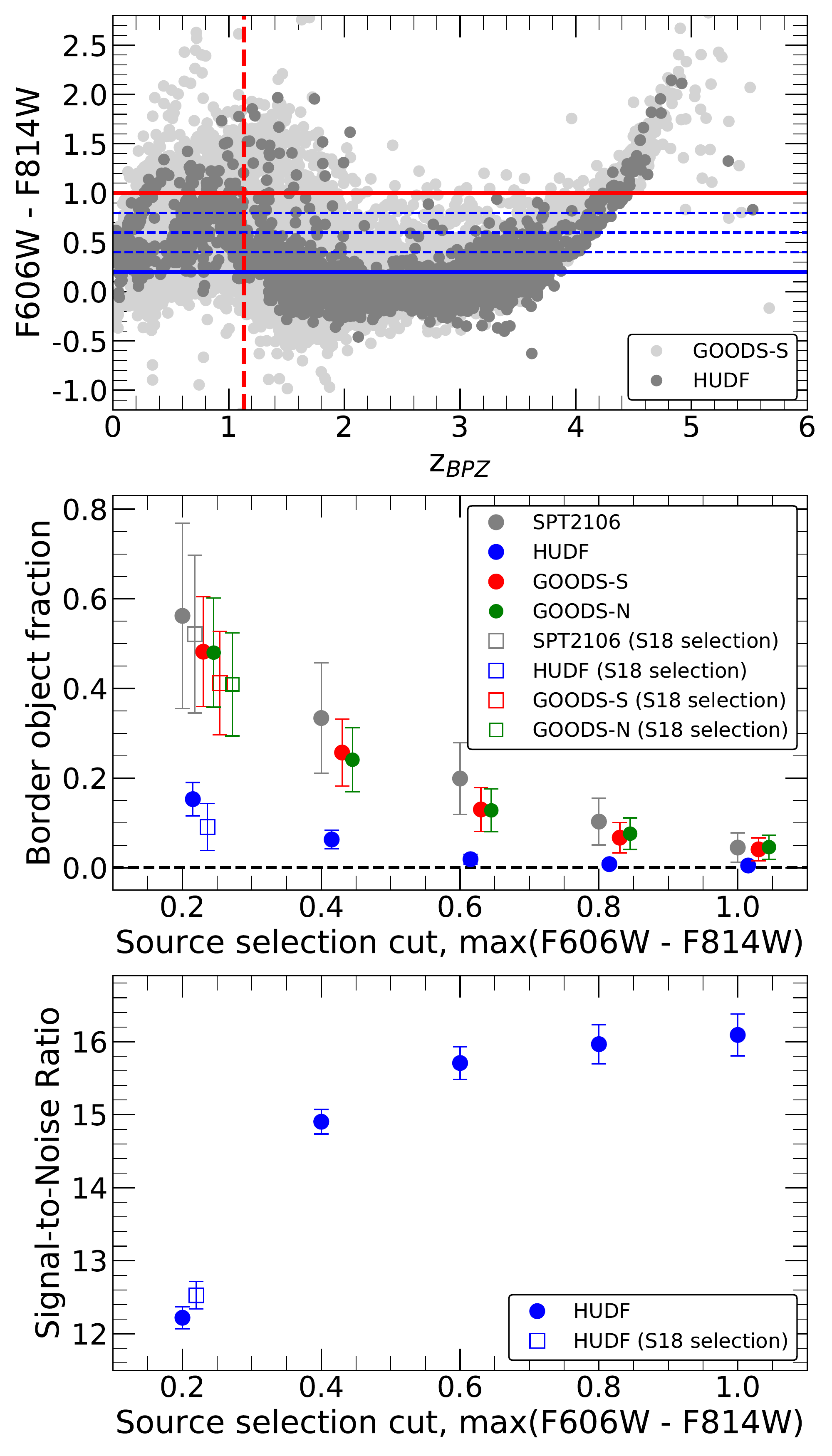}
\caption{Source selection color criteria versus S/N in WL mass measurement.
Top: color-redshift relation of the GOODS-S and HUDF galaxies within the magnitude range $25.0 < \mbox{F606W} < 28.0$. We use the published photo-$z$ catalogs. Red solid line represents our color threshold ($\mbox{F606W}-\mbox{F814W}< 1.0$) while blue solid line shows the \cite{Schrabback2018} threshold ($\mbox{F606W}-\mbox{F814W}< 0.2$).
Three blue dashed lines indicate our experimental color thresholds whose impacts are shown in the middle and bottom panels.
The vertical red dashed line is the redshift of SPT2106~($z=1.132$). 
Note that the S18 threshold is motivated by the tight color-redshift relation in HUDF. Apparently, the relation in GOODS-S is significantly diluted.\\
Middle: fraction of the ``border" objects (see text for the definition) as a function of color threshold. To avoid clutters, arbitrary offsets are added to the horizontal axis.
The fraction of ``border" sources increases as the $\mbox{F606W}-\mbox{F814W}$ color threshold decreases. This effect is more significant if the field is shallower. 
We also display the results when we apply the magnitude cut ($24.0 < \mbox{F606W} < 26.5$) of S18 (open squares). \\
Bottom: S/N in WL mass as a function of color threshold. The result shows the combined impact of purity, shot noise, and $\left < \beta \right >$ uncertainty due to the sample variance (see text).
We estimate the errors in S/N computation assuming Poissonian statistics.
Although lowering the $\mbox{F606W}-\mbox{F814W}$ color threshold may increase the purity of background sources, the increase in shot noise and $\left < \beta \right >$ uncertainty causes the total S/N to decrease.
}
\label{fig:color_redshift}
\end{figure}

\subsubsection{Source selection and signal-to-noise ratio}
The two WL studies use different source selection (especially color cut) criteria;
S18 select bluer ($\mbox{F606W}-\mbox{F814W}< 0.2$) and brighter ($24.0 < \mbox{F606W} < 26.5$) sources while we include redder ($\mbox{F606W}-\mbox{F814W}< 1.0$) and fainter ($25.0 < \mbox{F606W} < 28.0$) objects.
S18 claim that their color cut maximizes the background galaxy fraction and the expected (non-background) contamination rate is $\mytilde2.4$\%.

When the color-redshift relation is tight and the photometric scatter is negligible, we agree with S18 that the S18 color cut is optimal for maximizing the background fraction.
As shown in the top panel of Figure~\ref{fig:color_redshift}, these conditions are satisfied
for the HUDF galaxies. Most sources bluer than $\mbox{F606W}-\mbox{F814W}< 0.2$ are above the cluster redshift $z=1.132$ (dark gray).
However, this color-redshift relation is substantially diluted if the field becomes shallower. Even for the GOODS-S field, which is deeper than our cluster field, the scatter is severe (light gray). 

Given the large increase in scatter for the color-redshift relation, we do not think that the $\mbox{F606W}-\mbox{F814W}< 0.2$ cut is still optimal for the cluster field. To investigate the issue quantitatively, we define ``border" objects as sources whose 1-$\sigma$ photometric error bars touch the color thresholds. A ``border" source can be included or excluded depending on the direction of the scatter. 
The fraction of ``border" objects increases as the image depth becomes shallower because of the increase in photometric error. The middle panel of Figure~\ref{fig:color_redshift} shows this trend for several choices of the $\mbox{F606W}-\mbox{F814W}$ color constraint. 
In the case of the bluest selection ($\mbox{F606W}-\mbox{F814W}< 0.2$), the border fraction is nearly $\mytilde50$\% while the fraction is negligible for our case ($\mbox{F606W}-\mbox{F814W}< 1$). Consequently, we believe that the purity (background fraction) is compromised non-negligible by the increased ``border" fraction when one uses a very tight color-redshift relation only observed in extremely deep fields such as HUDF.

Apart from the decrease in the border fraction mentioned above, our selection has two additional advantages: shot noise reduction and improved accuracy in $\left< \beta \right>$ estimation.
Our source density is more than a factor of 9 higher than the S18 one, which results in more than a factor of 3 decrease in shot noise. Also, the increase in source galaxy number leads to a significant reduction in the $\left< \beta \right>$ value uncertainty because of the decrease in sample variance between the cluster and control fields. 

Therefore, including red galaxies ($\mbox{F606W}-\mbox{F814W}< 1$) in our source selection enables us to increase the overall S/N in cluster mass estimation compared to the case where only blue sources are selected ($\mbox{F606W}-\mbox{F814W}< 0.2$). To illustrate this point, we display the mass estimation S/N value as a function of the color cut in the bottom panel of Figure~\ref{fig:color_redshift}. Although the purity of the background galaxies (i.e., numerator in S/N) decreases as the selection includes redder galaxies, the overall S/N value increases because of the reduction in the denominator comprised of the shot noise and the $\left< \beta \right>$ uncertainty.

\subsection{Rarity} \label{how_rare}
SPT2106 has been regarded as a rare system within the current \LCDM hierarchical structure formation paradigm (e.g., F11; \citealt{Holz2012}). 
Here we revisit the issue using our WL mass. Previous rarity analyses of this system have been based on non WL results.

The expected abundance of a cluster with the mass $M$ and the redshift $z$ over the sky coverage fraction $f_{sky}$
can be evaluated using the following integral \citep{Hotchkiss2011}:
\begin{equation}
N(M,z) = f_{sky} \int_{z_{min}}^{z_{max}} \frac{dV(z)}{dz} dz \int_{M_{min}}^{M_{max}} \frac{dn}{dM} 
dM 
\label{eqn_abundance}
\end{equation}
\noindent
where $dV/dz$ is the comoving volume element and $dn/dM$ is the halo mass function.
In order to compute the mass function, we use the open-source python package called {\tt HMFcalc} \citep{HMFcalc}.
We adopt the \cite{Tinker2008} fitting function and the \cite{Planck2016} cosmology.
The redshift of the cluster $z=1.132$ is chosen to be the minimum redshift and $M_{200c}$ is used for the abundance estimation. 

We evaluate the expected number of SPT2106-like clusters and their corresponding discovery probabilities assuming Poisson statistics in the parent survey volume (SPT-SZ survey, $2500$ sq. deg). 
For the threshold mass $M_{min}$, we use the $1\sigma$ lower limit of our result $M_{200c} = 10.4^{+3.3}_{-3.0}~\times$~\solarm.
With these conditions, we estimate the expected abundance to be $\mytilde1.3$, which yields the discovery probability $\mytilde72$\% for the parent survey volume. 

\cite{Mortonson2011} argue that Eddington bias \citep{Eddington1913} is an important factor affecting a discovery
probability when the mass function is steep, which makes up-scattering events more frequent than down-scattering events.
Following the \cite{Mortonson2011} prescription, we obtain the Eddington bias-corrected abundance $\mytilde2.8$ in the parent $2500$ sq. deg survey area.

Our rarity analysis shows that SPT2106 is certainly rare, but should not be regarded as an outlier within the current cosmology. 
In some previous studies (e.g., F11; \citealt{Holz2012}), the existence of SPT2106 was viewed as an extremely rare event ($\mytilde7$\% within the SPT survey volume). This is because the mass estimate of F11 is $\mytilde24$\% larger than our WL result, which causes a non-negligible difference in abundance estimation when the mass function is steep.

\subsection{Systematic Uncertainties} \label{systematic_errors}
The errors quoted in Table~\ref{table_2PMD_MCMC} and \ref{table_1PMD} include statistical errors only. 
Below, we discuss possible sources of systematics for our WL analysis and quantify their contributions to the total error budget. 
In summary, by adding in quadrature, 
we estimate that the total systematic uncertainty of our mass is $\mytilde9$\% at $r = 840$~kpc.

\subsubsection{Impacts of the sample variance on source redshift estimation} \label{sample_variance}
If found to be large, the sample variance 
weakens the reliability of our source redshift determination.
In \S\ref{section_beta}, we obtain the source redshift information using the HUDF photo-$z$ catalog by assuming that
the color-magnitude-redshift relation is identical between the cluster field and HUDF.
Because both fields are of similar size, we are interested in quantifying how much
the source redshift varies when a different sky patch is chosen.
Our previous studies \citep[e.g.,][]{Jee2014,Jee2017} have shown that in fact the sample variance is subdominant compared to the statistical errors. Here we repeat the same test to quantify the level of the sample variance on the SPT2106 field by comparing $\left < \beta \right >$ measured from the HUDF, GOODS-S, and GOODS-N control fields. 

We examine the sample variance on two angular scales. The first angular scale is the size of the GOODS fields. The second angular scale is the distance between GOODS-N and GOODS-S.
For the sample variance estimation on the first scale, we divide each GOODS field into 9-11 subregions 
and measure the fluctuation of the $\left < \beta \right >$ values. For ACS sources ($\mbox{F606W}-\mbox{F814W}$), the standard deviation of the distribution in $\left < \beta \right >$ is $\mytilde0.01$ for each GOODS field. For WFC3 sources ($\mbox{F105W}-\mbox{F140W}$), the distribution width decreases by a factor of two ($\sigma_{ \left < \beta \right >}\sim0.005$). We attribute this reduction to the greater depth of the WFC3/IR data, 
which gives a higher total volume because of its long line-of-sight baseline. 
These $\left < \beta \right >$ fluctuations are translated to less than $\mytilde7$\% in mass, which shows that the sample variance is negligible on the first angular scale (corresponding to the GOODS field size). For both ACS and WFC3 sources, the difference in the average $\left < \beta \right >$ value between GOODS-N and GOODS-S is $\mytilde0.01$, similar to the level of the fluctuations observed within each GOODS field. Although future studies still need to investigate the sample variance on the above and other angular scales with more than just these two fields, we have not found any strong evidence that the sample variance is a significant source of systematics when WL sources are selected from deep imaging data (sufficient volume along the-line-of-sight direction) and their redshift estimation is based on deep control fields.

\subsubsection{Systematics due to mass-concentration relations} \label{Mc_relations}
The use of mass-concentration relations can be a source of systematic uncertainties. 
In \S\ref{mass_estimate} we measure the mass estimates of SPT2106~with and without the $M-c$ relations. 
Figure~\ref{fig:MCMC_figs} shows the distributions of posteriors on the mass-concentration plane and their marginalized values. Also displayed are the mass estimates for three $M-c$ relations. 
The masses obtained from the three $M-c$ relations are consistent with the results derived from our two-parameter model. On the other hand, the two concentration values favored by the DM14 and DK15 $M-c$ relations are outside the $1\sigma$ contours.

Although applying an $M-c$ relation is an efficient and popular method to determine a cluster mass, we argue that
the mass obtained without the assumption is least biased when the relation is highly uncertain as in the current case.
Because massive high-$z$ clusters are rare even in the current large state-of-the-art $N$-body simulations, statistical
properties such as mass function, $M-c$ relations, etc. are only loosely determined.
For example, the $M-c$ relations for massive, high-$z$ clusters come from the power laws extrapolated from the results for less massive clusters; note that the shapes of the two power laws in \cite{DM14} and \cite{DK15} are different (e.g., flattening, upturn, etc.) in the high-redshift regime. 
In addition, the $M-c$ relations are valid for the ``mean" (or ``median") properties of simulated clusters. Thus there is no guarantee that SPT2106 follows the mean $M-c$ relation even if the extrapolated relation holds.

\begin{figure}
\centering
\includegraphics[width=80mm]{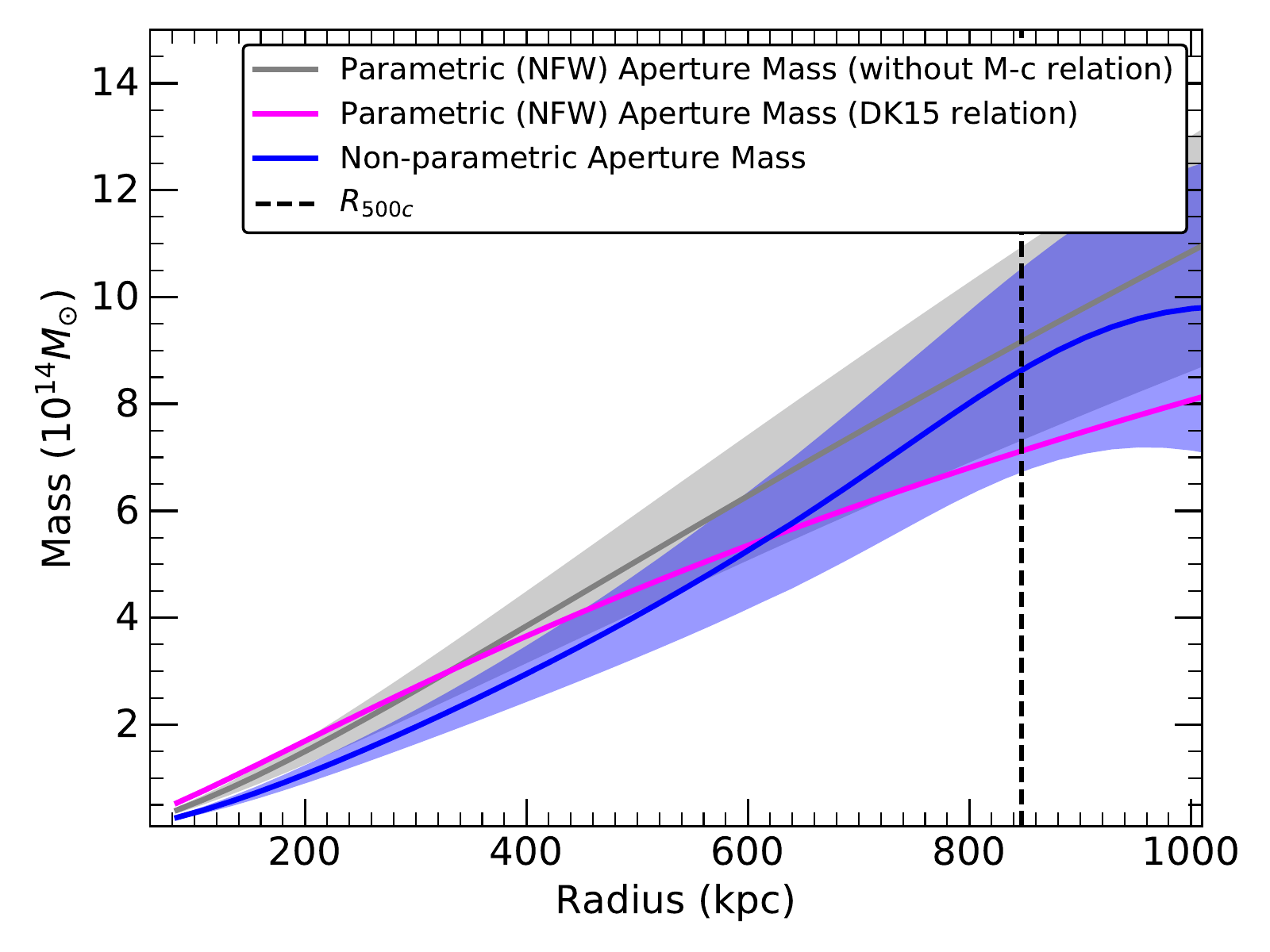}
\caption{
Projected mass of SPT2106. We evaluate the projected mass as a function of radius, the distance from the WL mass peak. The gray solid line represents the result after we project the best-fit NFW model (Table~\ref{table_2PMD_MCMC}). The blue solid line is obtained when we use the aperture mass densitometry (see text). Shaded regions depict 1-$\sigma$ uncertainty intervals.
Both results are highly consistent with each other. 
At $R_{500c}\sim840$~kpc, the difference is only $\mytilde6$\%, much less than the statistical uncertainties ($\mytilde22$\%). 
The agreement shows that the potential model bias due to our NFW assumption is not a major source of systematics.
}
\label{fig:AMD}
\end{figure}

\subsubsection{Cluster Model Bias} \label{model_bias}
In \S\ref{mass_estimate}, we assume a spherical NFW profile for measuring the mass of SPT2106. 
This assumption, however, is only applicable to a cluster with average properties. It is possible that SPT2106 
might deviate greatly from this mean cluster profile, which if true is an additional source of uncertainties in our mass determination. According to \cite{Becker2011}, this cluster model bias amounts to $\mytilde20$\% for massive clusters.
Some impacts of the assumed profile on our cluster mass estimation can be seen in Table~\ref{table_1PMD}, where we compare mass estimates using SIS and NFW profiles with various $M-c$ relations.
For example, the SIS mass estimate is $\mytilde30$\% lower than our two-parameter NFW result; we already discussed the impacts of the various $M-c$ relations in \S\ref{Mc_relations}.

As an attempt to assess the amount of systematics due to the cluster model bias, we perform aperture mass densitometry (hereafter AMD), 
which minimizes the impact from the assumption on any particular halo model. 
AMD uses tangential shears of sources (Equation~\ref{tan_shear}) and computes a projected overdensity of the region inside a specific aperture radius with respect to the control annulus. For the estimation of the mean surface density in the control annulus, we inevitably have to assume a particular halo model. Nevertheless, the assumption plays a much less significant role than in the pure model fitting method.

The projected overdensity within the $r=r_1$ aperture is given by: 
\begin{eqnarray}
\zeta_c (r_1, r_2,r_{max}) =  \bar{\kappa}( r \leq r_1) -
\bar{\kappa}( r_2 < r \leq r_{max}) \nonumber \\ = 2 \int_{r_1}^{r_2} \frac{
  \left < \gamma_T \right > }{r}dr + \frac{2}{1-r_2^2/r_{max}^2}
\int_{r_2}^{r_{max}} \frac{ \left <\gamma_T \right >}{r} dr,
\label{eqn_AMD}
\end{eqnarray}
\noindent
where $\left < \gamma_T \right>$ is the azimuthal average of tangential shears, $r_1$ is the aperture radius, and $r_2$ and $r_{max}$ are the inner and the outer radii of the annulus, respectively. 
The aperture radius should be sufficiently separated from the inner radius of the annulus because the AMD uses tangential shears between the aperture radius $r_1$ and the inner radius $r_2$ for the integral evaluation as expressed in Equation~\ref{eqn_AMD}. 
The statistics $\zeta_c(r_1,r_2,r_{max})$ is then converted into the aperture mass within the radius $r_1$ through the equation below:
\begin{equation}
M_{proj} (r < r_1) = \pi r_1^2 \zeta_c(r_1) \Sigma_{crit}, \label{eqn_AMD_mass}
\end{equation}
\noindent
where $\Sigma_{crit}$ is the critical surface density obtained in Equation~\ref{eqn_sigma_c}.

We choose the inner and outer radii of the control annulus to be $155$\arcsec~($\mytilde1.27$~Mpc) and $175$\arcsec~($\mytilde1.44$~Mpc), respectively. 
The average surface density of the annulus is estimated to be $\bar{\kappa} = 0.009$ from our two-parameter NFW fitting result (Table~\ref{table_2PMD_MCMC}). 
We iteratively update the tangential shear using $\gamma = (1 - \kappa) g$ because what we measure is not a strict shear $\gamma$ but a reduced shear $g=\gamma/(1-\kappa)$. 
The process stops when the density is converged. 
Readers are referred to \cite{Clowe2000} and \cite{Jee2005} for details of the AMD implementation.

Figure~\ref{fig:AMD} shows the resulting cumulative projected mass profile. Also displayed are the profiles derived from the best-fit NFW models for comparison. 
At $r=840$~kpc, the AMD mass agrees with our NFW result obtained with two-parameter fitting within $\mytilde6$\%, which we adopt as the systematic error due to the model bias. The shaded region represents the $1\sigma$ uncertainties computed from bootstrapping runs. The statistical error at the radius is $\mytilde22$\%. 
We find a $\mytilde21$\% difference at the same radius when we compare the AMD result with the NFW result based on the DK15 $M-c$ relation. 
These comparisons demonstrate that an NFW profile assumption might not be a significant source of systematics in our determination of the SPT2106 mass. And the excellent agreement with the AMD profile shows that the intrinsic mass profile of SPT2106 might be better described with the NFW parameters derived from our two-parameter fitting, although the result from the DK15 $M-c$ relation is still marginally consistent with the AMD profile.

The good agreement of the AMD profile with our parametric (NFW) result also suggests that the mass profile of SPT2106 can be well approximated by a single halo. In general, treating well-separated merging clusters as a single halo is an important source of bias in total mass estimation of the systems (e.g., \citealt{Jee2014}). However, since in SPT2106 the two substructures is very close ($\mytilde150$~kpc), we believe that this single-halo approximation is not a significant source systematic error.

\section{Summary} \label{section_summary}
We have presented a detailed WL study of the massive high-$z$ cluster \spt~based on new \HST~data.
From the deep \HST~imaging data, we achieve a source density of 169 $\mbox{arcmin}^{-2}$ in the WFC3/IR field, which enables us to
perform a high-resolution mass reconstruction and improve the precision in mass estimation.

The overall mass distribution of the cluster is characterized by the main mass clump and the extension $\mytilde0.6$~Mpc west of it. These two structures are spatially correlated with the cluster galaxy distributions. 
The main clump is further resolved into two substructures. The northwestern mass peak position agrees with that of the X-ray peak, which also coincides with the BCG whereas the southeastern peak seems to be associated with the number density peak. 
This WL substructure supports the merger hypothesis suggested in previous studies.

We estimate the cluster mass with various assumptions on its profile and argue that the result obtained with the two-parameter NFW profile fitting (i.e., without any assumption on the $M-c$ relation) method is
likely to be the least biased mass because the $M-c$ relation is highly uncertain for extremely massive clusters at high redshift. We support this claim with our parameter-free result based on aperture mass densitometry. Our virial mass $M_{200c} = (10.4^{+3.3}_{-3.0}\pm1.0)~\times$~\solarm~(the second error bar is the systematic uncertainty) agrees nicely
with the previous measurements obtained from SZ and X-ray studies.
Our statistical uncertainty is $\mytilde40$\% smaller than the previous WL result. The difference is attributed to a significant source density
increase due to the availability of the deep WFC3 imaging data and improvement in shape measurement.
We find that given our new WL mass estimate, the cluster SPT2106 is certainly rare, but should not be regarded as an outlier within the current \LCDM cosmology.

\acknowledgments
Support for the current \HST~program was provided by NASA through a grant from the Space Telescope Science
Institute, which is operated by the Association of Universities for Research in Astronomy, Incorporated, under
NASA contract NAS5-26555.
We thank Kyle Finner and Cristiano Sabiu for useful discussions.
M. J. Jee acknowledges support for the current research from the National Research Foundation (NRF) of Korea under the programs 2017R1A2B2004644 and 2017R1A4A1015178. 
J. Kim acknowledges support from Shin Donguk scholarship at Yonsei University 
and the NRF of Korea grant funded by the Ministry of Science and ICT, Korea under the program 2019R1C1C1010942.

\bibliographystyle{aasjournal}

\begin{thebibliography}{}
\bibitem[Adami et al.(2010)]{Adami2010} Adami, C., Durret, F., Benoist, C., et al.\ 2010, \aap, 509, A81
\bibitem[Allen et al.(2011)]{Allen2011} Allen, S.~W., Evrard, A.~E., \& Mantz, A.~B.\ 2011, \araa, 49, 409
\bibitem[Amodeo et al.(2016)]{Amodeo2016} Amodeo, S., Ettori, S., Capasso, R., \& Sereno, M.\ 2016, \aap, 590, A126
\bibitem[Bartalucci et al.(2017)]{Bartalucci2017} Bartalucci, I., Arnaud, M., Pratt, G.~W., et al.\ 2017, \aap, 598, A61
\bibitem[Becker \& Kravtsov(2011)]{Becker2011} Becker, M.~R., \& Kravtsov, A.~V.\ 2011, \apj, 740, 25
\bibitem[Beckwith et al.(2006)]{UDF06} Beckwith, S.~V.~W., Stiavelli, M., Koekemoer, A.~M., et al.\ 2006, \aj, 132, 1729
\bibitem[Bertin \& Arnouts(1996)]{Bertin1996} Bertin, E., \& Arnouts, S.\ 1996, \aaps, 117, 393
\bibitem[Bleem et al.(2015)]{Bleem2015} Bleem, L.~E., Stalder, B., de Haan, T., et al.\ 2015, \apjs, 216, 27
\bibitem[Chiu et al.(2016)]{Chiu2016} Chiu, I., Mohr, J., McDonald, M., et al.\ 2016, \mnras, 455, 258
\bibitem[Clowe et al.(2000)]{Clowe2000} Clowe, D., Luppino, G.~A., Kaiser, N., \& Gioia, I.~M.\ 2000, \apj, 539, 540 
\bibitem[Diemer \& Kravtsov(2015)]{DK15} Diemer, B., \& Kravtsov, A.~V.\ 2015, \apj, 799, 108 
\bibitem[Duffy et al.(2008)]{Duffy08} Duffy, A.~R., Schaye, J., Kay, S.~T., \& Dalla Vecchia, C.\ 2008, \mnras, 390, L64 
\bibitem[Dutton \& Macci{\`o}(2014)]{DM14} Dutton, A.~A., \& Macci{\`o}, A.~V.\ 2014, \mnras, 441, 3359 
\bibitem[Eddington(1913)]{Eddington1913} Eddington, A.~S.\ 1913, \mnras, 73, 359
\bibitem[Erben et al.(2001)]{Erben2001} Erben, T., Van Waerbeke, L., Bertin, E., Mellier, Y., \& Schneider, P.\ 2001, \aap, 366, 717
\bibitem[Fassbender et al.(2011)]{Fassbender2011} Fassbender, R., Nastasi, A., B{\"o}hringer, H., et al.\ 2011, \aap, 527, L10
\bibitem[Fischer \& Tyson(1997)]{FIATMAP} Fischer, P., \& Tyson, J.~A.\ 1997, \aj, 114, 14
\bibitem[Foley et al.(2011)]{Foley2011} Foley, R.~J., Andersson, K., Bazin, G., et al.\ 2011, \apj, 731, 86
\bibitem[Foreman-Mackey et al.(2013)]{emcee} Foreman-Mackey, D., Hogg, D.~W., Lang, D., \& Goodman, J.\ 2013, \pasp, 125, 306 
\bibitem[Fruscione et al.(2006)]{CIAO} Fruscione, A., McDowell, J.~C., Allen, G.~E., et al.\ 2006, Society of Photo-optical Instrumentation Engineers (SPIE) Conference Series, 62701V
\bibitem[Giavalisco et al.(2004)]{GOODSmag} Giavalisco, M., Ferguson, H.~C., Koekemoer, A.~M., et al.\ 2004, \apjl, 600, L93
\bibitem[Golovich et al.(2017)]{Golovich2017} Golovich, N., van Weeren, R.~J., Dawson, W.~A., Jee, M.~J., \& Wittman, D.\ 2017, \apj, 838, 110 
\bibitem[Hack et al.(2003)]{CALACS} Hack, W., Busko, I., \& Jedrzejewski, R.\ 2003, ASP Conf.~Ser.~295: Astronomical Data Analysis Software and Systems XII, 295, 453
\bibitem[Hoekstra et al.(1998)]{Hoekstra1998} Hoekstra, H., Franx, M., Kuijken, K., \& Squires, G.\ 1998, \apj, 504, 636
\bibitem[Holz \& Perlmutter(2012)]{Holz2012} Holz, D.~E., \& Perlmutter, S.\ 2012, \apjl, 755, L36 
\bibitem[Hotchkiss(2011)]{Hotchkiss2011} Hotchkiss, S.\ 2011, \jcap, 7, 004
\bibitem[Jee et al.(2005)]{Jee2005} Jee, M.~J., White, R.~L., Ben{\'{\i}}tez, N., et al.\ 2005, \apj, 618, 46
\bibitem[Jee et al.(2007)]{Jee2007} Jee, M.~J., Blakeslee, J.~P., Sirianni, M., et al.\ 2007, \pasp, 119, 1403
\bibitem[Jee \& Tyson(2009)]{JeeTyson2009} Jee, M.~J., \& Tyson, J.~A.\ 2009, \apj, 691, 1337 
\bibitem[Jee \& Tyson(2011)]{JeeTyson2011} Jee, M.~J., \& Tyson, J.~A.\ 2011, \pasp, 123, 596
\bibitem[Jee et al.(2011)]{Jee2011} Jee, M.~J., Dawson, K.~S., Hoekstra, H., et al.\ 2011, \apj, 737, 59
\bibitem[Jee et al.(2013)]{Jee2013} Jee, M.~J., Tyson, J.~A., Schneider, M.~D., et al.\ 2013, \apj, 765, 74
\bibitem[Jee et al.(2014)]{Jee2014} Jee, M.~J., Hughes, J.~P., Menanteau, F., et al.\ 2014, \apj, 785, 20 
\bibitem[Jee et al.(2017)]{Jee2017} Jee, M.~J., Ko, J., Perlmutter, S., et al.\ 2017, \apj, 847, 117 
\bibitem[Kaiser \& Squires(1993)]{KS93} Kaiser, N., \& Squires, G.\ 1993, \apj, 404, 441
\bibitem[Kaiser et al.(1995)]{KSB1995} Kaiser, N., Squires, G., \& Broadhurst, T.\ 1995, \apj, 449, 460 
\bibitem[Koekemoer et al.(2002)]{2002multidrizzle} Koekemoer, A. M., Fruchter, A. S., Hook, R. N., \& Hack, W. 2002, in The 2002 HST Calibration Workshop, ed. S. Arribas, A. Koekemoer, \& B. Whitmore (Baltimore: STScI), 337
\bibitem[Kim et al.(2019)]{Mincheol2019} Kim, M., Jee, M.~J., Finner, K., et al.\ 2019, \apj, 874, 143
\bibitem[Mandelbaum et al.(2015)]{GREAT3} Mandelbaum, R., Rowe, B., Armstrong, R., et al.\ 2015, \mnras, 450, 2963
\bibitem[Markwardt(2009)]{MPFIT} Markwardt, C.~B.\ 2009, Astronomical Data Analysis Software and Systems XVIII, 411, 251
\bibitem[Marriage et al.(2011)]{Marriage2011} Marriage, T.~A., Acquaviva, V., Ade, P.~A.~R., et al.\ 2011, \apj, 737, 61
\bibitem[Massey et al.(2014)]{Massey2014} Massey, R., Schrabback, T., Cordes, O., et al.\ 2014, \mnras, 439, 887
\bibitem[Mehrtens et al.(2012)]{Mehrtens2012} Mehrtens, N., Romer, A.~K., Hilton, M., et al.\ 2012, \mnras, 423, 1024
\bibitem[Melchior \& Viola(2012)]{Melchior2012} Melchior, P., \& Viola, M.\ 2012, \mnras, 424, 2757
\bibitem[Mortonson et al.(2011)]{Mortonson2011} Mortonson, M.~J., Hu, W., \& Huterer, D.\ 2011, \prd, 83, 023015
\bibitem[Murray et al.(2013)]{HMFcalc} Murray, S.~G., Power, C., \& Robotham, A.~S.~G.\ 2013, Astronomy and Computing, 3, 23
\bibitem[Muzzin et al.(2009)]{Muzzin2009} Muzzin, A., Wilson, G., Yee, H.~K.~C., et al.\ 2009, \apj, 698, 1934
\bibitem[Navarro et al.(1997)]{NFW1997} Navarro, J.~F., Frenk, C.~S., \& White, S.~D.~M.\ 1997, \apj, 490, 493
\bibitem[Planck Collaboration et al.(2016)]{Planck2016} Planck Collaboration, Ade, P.~A.~R., Aghanim, N., et al.\ 2016, \aap, 594, A13
\bibitem[Rafelski et al.(2015)]{UVUDF} Rafelski, M., Teplitz, H.~I., Gardner, J.~P., et al.\ 2015, \aj, 150, 31
\bibitem[Refregier et al.(2012)]{Refregier2012} Refregier, A., Kacprzak, T., Amara, A., Bridle, S., \& Rowe, B.\ 2012, \mnras, 425, 1951 
\bibitem[Reichardt et al.(2013)]{Reichardt2013} Reichardt, C.~L., Stalder, B., Bleem, L.~E., et al.\ 2013, \apj, 763, 127
\bibitem[Rowe et al.(2015)]{Galsim} Rowe, B.~T.~P., Jarvis, M., Mandelbaum, R., et al.\ 2015, Astronomy and Computing, 10, 121
\bibitem[Ruel et al.(2014)]{Ruel2014} Ruel, J., Bazin, G., Bayliss, M., et al.\ 2014, \apj, 792, 45
\bibitem[Schrabback et al.(2007)]{Schrabback2007} Schrabback, T., Erben, T., Simon, P., et al.\ 2007, \aap, 468, 823
\bibitem[Schrabback et al.(2010)]{Schrabback2010} Schrabback, T., Hartlap, J., Joachimi, B., et al.\ 2010, \aap, 516, A63
\bibitem[Schrabback et al.(2018)]{Schrabback2018} Schrabback, T., Applegate, D., Dietrich, J.~P., et al.\ 2018, \mnras, 474, 2635
\bibitem[Seitz \& Schneider(1997)]{Seitz1997} Seitz, C., \& Schneider, P.\ 1997, \aap, 318, 687
\bibitem[Skelton et al.(2014)]{CANDELS} Skelton, R.~E., Whitaker, K.~E., Momcheva, I.~G., et al.\ 2014, \apjs, 214, 24
\bibitem[Song et al.(2012)]{Song2012} Song, J., Zenteno, A., Stalder, B., et al.\ 2012, \apj, 761, 22
\bibitem[Tinker et al.(2008)]{Tinker2008} Tinker, J., Kravtsov, A.~V., Klypin, A., et al.\ 2008, \apj, 688, 709
\bibitem[Ubeda \& Anderson(2012)]{Ubeda2012} Ubeda, L., \& Anderson, J.\ 2012,Instrument Science Report, ACS 2012-03 (Baltimore: STScI)
\bibitem[van Weeren et al.(2017)]{vanWeeren2017} van Weeren, R.~J., Andrade-Santos, F., Dawson, W.~A., et al.\ 2017, Nature Astronomy, 1, 0005
\bibitem[Williamson et al.(2011)]{Williamson2011} Williamson, R., Benson, B.~A., High, F.~W., et al.\ 2011, \apj, 738, 139

\end{thebibliography}

\clearpage

\end{document}